\documentclass[prb,twocolumn,superscriptaddress,10pt,
floatfix
]{revtex4-1}

\usepackage{amsmath}
\usepackage{amssymb}
\usepackage{graphicx}
\usepackage{bm}
\usepackage[top=3cm, bottom=3.25cm, left=2.75cm, right=2.75cm]{geometry}
\usepackage{mathrsfs}
\usepackage{subfigure}
\usepackage{booktabs}
\usepackage{siunitx}
\usepackage{makecell, booktabs} 
\usepackage{footnote}
\usepackage{xcolor}

\def\be{\begin{equation}}
\def\ee{\end{equation}}
\def\bea{\begin{eqnarray}}

\def\bea{\end{eqnarray}}
\def\eea{\end{eqnarray}} 	
\def\bs{\begin{split}}
\def\es{\end{split}}
\def\ni{\noindent}

\def\bi{\begin{itemize}}
\def\ei{\end{itemize}}

\def\m{\mu}

\def\s{\sigma}

\def\f{\frac}

\begin{document}

\title{\Large{Ab-initio energetics of graphite and multilayer graphene: \\ stability of Bernal versus rhombohedral stacking}}
\vspace{3 mm}
\author{Jean Paul Nery}
\email{jeanpaul240@gmail.com}
\affiliation{Graphene Labs, Fondazione Istituto Italiano di Tecnologia, Via Morego, I-16163 Genova, Italy.}
\affiliation{Dipartimento di Fisica, Universit\`a di Roma La Sapienza, Piazzale Aldo Moro 5, I-00185 Roma, Italy.}
\author{Matteo Calandra}
\affiliation{Graphene Labs, Fondazione Istituto Italiano di Tecnologia, Via Morego, I-16163 Genova, Italy.}
\affiliation{Sorbonne Universit\'e, CNRS, Institut des Nanosciences de Paris, UMR7588, F-75252, Paris, France.}
\affiliation{Department of Physics, University of Trento, Via Sommarive 14, 38123 Povo, Italy.}
\author{Francesco Mauri}
\email{francesco.mauri@uniroma1.it}
\affiliation{Graphene Labs, Fondazione Istituto Italiano di Tecnologia, Via Morego, I-16163 Genova, Italy.}
\affiliation{Dipartimento di Fisica, Universit\`a di Roma La Sapienza, Piazzale Aldo Moro 5, I-00185 Roma, Italy.}


\begin{abstract} There has been a lot of excitement around the observation of superconductivity in twisted bilayer graphene, associated to flat bands close to the Fermi level. Such correlated electronic states also occur in multilayer rhombohedral stacked graphene (RG), which has been receiving increasing attention in the last years. In both natural and artificial samples however, multilayer stacked Bernal graphene (BG) occurs more frequently, making it desirable to determine what is their relative stability and under which conditions RG might be favored. Here, we study the energetics of BG and RG in bulk and also multilayer stacked graphene using first-principles calculations. It is shown that the electronic temperature, not accounted for in previous studies, plays a crucial role in determining which phase is preferred. We also show that the low energy states at room temperature consist of BG, RG and mixed BG-RG systems with a particular type of interface. Energies of all stacking sequences (SSs) are calculated for $N=12$ layers, and an Ising model is used to fit them, which can be used for larger $N$ as well. In this way, the ordering of low energy SSs can be determined and analyzed in terms of a few parameters. Our work clarifies inconsistent results in the literature, and sets the basis to studying the effect of external factors on the stability of multilayer graphene systems in first principles calculations.
\end{abstract}
\maketitle

\section{Introduction}

Multilayer graphene has in principle many possible stable configurations. If the lower two layers are fixed in configuration AB, $N$ layers of graphene have $2^{N-2}$ local minima, since each additional layer can take two possible positions. Of all these possible stacking sequences (SSs), the two that are usually observed exhibit periodicity. The most common one has AB stacking and is known as Bernal-stacked multilayer graphene (BG). The other one is rhombohedral-stacked multilayer graphene (RG), which has ABC stacking, but its occurence is less frequent\cite{Lipson1942}. Our interest lies in RG, due to the presence of flat bands close to the Fermi level \cite{Pierucci2015,Henck2018,Koshino2010,Xiao2011}, which makes it a candidate for exciting phenomena like high temperature superconductivity\cite{Kopnin2013,Munioz2013}, and also charge-density wave or magnetic states \cite{Otani2010,Pamuk2017,Baima2018}. In fact, superconductivity might have already been observed in interfaces between BG and RG \cite{Precker2016,Ballestar2013}.

The region of the Brillouin zone where the bands are flat increases with the number of layers, up to about 8 layers \cite{Koshino2010,Pamuk2017}. It has been known for a long time, through X-ray diffraction measurements, that shear increases the percentage of RG\cite{Laves1956,Boehm1964}. Furthermore, we showed recently that shear stress can lead to many consecutive layers of RG \cite{Nery2020}. All of the samples where long-range rhombohedral order has been observed involved shear stress, due to milling\cite{Lin2012}, exfoliation \cite{Henni2016,Henck2018,Yang2019}, or steps in copper substrates used in chemical vapor deposition (CVD) in Ref.~\onlinecite{Chams2020}. In this last work, RG with a thickness of up to 9 layers and area of up to 50 $\m$m$^2$ was grown, while previous studies were limited to widths of 100 nm (see references within Ref.~\onlinecite{Chams2020}), which is important for device fabrication. Other approaches that could lead to long-range rhombohedral order (and large areas), but that so far have lead to only a few layers of RG, include curvature\cite{Gao2020}, twisting\cite{Kerelsky2019}, an electric field\cite{Yankowitz2014,Li2020}, and doping\cite{Li2020}. According to Ref.~\onlinecite{Geisenhof2019}, anisotropic strain favors BG, so it should be avoided during transfer to a substrate (like hexagonal boron nitride) or upon deposition of metal contacts. The authors also suggested removing all non-rhombohedral parts prior to transfer (via etching, or cutting with the tip of an atomic force microscope\cite{Chen2019}). This is in line with the observation of Ref.~\onlinecite{Yang2019} that, when transferring flakes with some rhombohedral order, such order was more likely to be preserved after transfer if the whole flake was ABC-stacked.

Despite the immense interest in graphene systems, the relative energies of different SSs are still not well established in general (neither experimentally nor theoretically), and in particular between BG and RG. Understanding the precise external conditions under which RG is favored is fundamental to better control its production and preservation. To achieve this, it would be useful to first determine the energy of different SSs without any such external factors (such as shear, strain, curvature, twisting, pressure, doping, or electric fields). We only know of 2 experimental works that determine the energy difference between RG and BG. More than 50 years ago, Ref.~\onlinecite{Boehm1964} estimated it by using a calorimeter and the enthalpy of formation of potassium graphite, and obtained 0.61 meV/atom. A more recent experiment\cite{Li2020} used small-angle twisted few layer graphene with different doping levels and electric fields. Through the curvature of stacking domain walls, energy differences between -0.05 and 0.10 meV/atom were obtained. The data was fit with a tight-binding model (for the energy dispersion of ABA and ABC) 
and used $E_\textrm{ABC}-E_\textrm{ABA}=0.05$ meV/atom. Density functional theory (DFT) calculations obtain similar values (although with both positive and negative signs, as we see in the first section in more detail). A related quantity is the stacking fault (SF) energy, which has been measured and calculated to be 0.14 meV/interface-atom (see Table~\ref{tab:LDA}). Thus, more recent works suggest that the energy differences between BG and RG are in the 0.01-0.10 meV/atom range, and that the value of Ref.~\onlinecite{Boehm1964} is overestimated.\\
\indent Here, we study the energy differences between BG and RG in bulk, and the energy distribution of all SSs for different number of layers $N$. Our calculations use DFT within the local density functional approximation (LDA). First, we calculate the bulk ($N \rightarrow \infty$) energy differences, and justify why they are so small. Then, it is shown that the low energy configurations at room temperature are probably BG, RG and mixed BG-RG phases with a particular type of interface which we refer to as soft. In addition, the stability of BG and RG is studied at different electronic temperatures and for different amount of layers, and a temperature vs. $N$ phase diagram is obtained.
Finally, we introduce a spin Ising model to fit the DFT energies, which helps to better understand how are SS distributed, and can also be used to predict the order of configurations for larger $N$.

\section{Bulk energy differences} 
\subsection{Different functionals and temperatures}

We first focus on the bulk energy difference between RG and BG, since there are contradicting results in the literature. DFT calculations have obtained $E_\textrm{RG}-E_\textrm{BG} \sim 0.1$ meV/atom \cite{Charlier1994,Savini2011,Anees2014} 
, but results are probably not well converged (for example, due to the use of a large smearing factor of 0.1 eV [\onlinecite{Anees2014}]). Ref.~\onlinecite{Taut2013} instead used very dense grids, and obtained $E_\textrm{RG}-E_\textrm{BG}<0$, which agrees with our room temperature result. However, the electronic temperature, which plays a fundamental role, was not considered.
 
The electronic temperature can be incorporated by using Fermi-Dirac smearing. We considered temperatures of 284 K (which we will refer to as room temperature), 1010 K and 3158 K (0.0018 Ry, 0.0064 Ry, and 0.02 Ry, respectively) and also several functionals (Table~\ref{tab:functionals}). As we can see, the difference is negative at room temperature (favoring RG), and positive at higher temperatures (favoring BG) for all functionals.
 Although these energy differences are small, the fact that all results are similar, and that the temperature trend is the same for all functionals, indicates that the results are meaningful. 
In Fig.~\ref{fig:bulk_vs_T}, we also plot the energy difference as a function of temperature, which shows BG becomes more stable at about 730 K.

In the rest of this work we will use an LDA functional, since it gives good energy differences. For example, the LDA shear frequency $\omega = 42$ cm$^{-1}$, related to the curvature of the energy landscape around the minimum AB, agrees well with experiment (see Ref.~\onlinecite{Nery2020} for more details).  Also, Table~\ref{tab:LDA} shows that it is in good agreement with more sophisticated methods like the adiabatic-connection fluctuation-dissipation theorem within the
random phase approximation (RPA) \cite{Wang2015,Zhou2015} and Quantum Monte Carlo (QMC)\cite{Mostaani2015}. The intermediate point between the minima AB and AC corresponds to a saddle point, and LDA agrees well with RPA. For AA stacking, LDA falls in between RPA and QMC. These two last values differ by 41\%, so more theoretical studies are needed to precisely determine energy differences in multilayer graphene systems.
\begin{table}[h]
{\begin{tabular}{cccc}
\hline \hline
\multicolumn{1}{l}{}  & \multicolumn{3}{c}{Energy difference (meV/atom)} \\
\multicolumn{1}{l}{}  & \begin{tabular}[c]{@{}c@{}} T=284K \hspace{2mm} \end{tabular} & \begin{tabular}[c]{@{}c@{}} T=1010K\end{tabular} \hspace{2mm}  & T=3158K\\
\hline
 LDA & -0.025         &  +0.020  & + 0.138 \\
\hline \\
Grimme  & -0.025	 &  +0.020  & +0.138 \\
\hline \\
GGA-PBE   & -0.040	 &  +0.002  & +0.116 \\
\hline \hline \\
\end{tabular}} \\
\caption{Bulk energy difference $E_\textrm{RG}-E_\textrm{BG}$ for different functionals at different electronic temperatures, using the experimental lattice parameter $a=2.461$ \AA$ $ and an interlayer distance $c_d=3.347$ \AA$ $. LDA is used throughout the rest of the paper. In Grimme\cite{Grimme2006}, an empirical $1/R_{ij}^6$ correction is added to the interaction between atoms, where $R_{ij}$ is the distance between them. The results are very similar to the LDA ones, which means that van der Waals corrections are not relevant in this semiempirical form. This is consistent with Ref.~\onlinecite{Taut2013}, where the Grimme correction is of only 0.5 $\m$eV. In the GGA-PBE case\cite{Perdew1996}, RG is even more favored at room temperature. All functionals give similar results at different temperatures.}
\label{tab:functionals}
\end{table}

\begin{table}[h]
{\begin{tabular}{ccc}
\hline \hline
\multicolumn{1}{l}{}  & \multicolumn{2}{c}{Energy (meV/atom)} \\
\multicolumn{1}{l}{}  & \begin{tabular}[c]{@{}c@{}} LDA \end{tabular}  & \begin{tabular}[c]{@{}c@{}} Previous works\end{tabular} \\
\hline
\begin{tabular}[c]{@{}c@{}} Stacking fault Bernal \\
ABABCACA \end{tabular} & 0.07                &  \begin{tabular}[c]{@{}c@{}} 0.14 (RPA) \cite{Wang2015}           \\ 0.14 (Exp.)\cite{Telling2003} \end{tabular} \\
\hline \\
\begin{tabular}[c]{@{}c@{}} Saddle point \\ (orthorombic) \end{tabular}   & 1.58  &  1.53 (RPA) \cite{Zhou2015}           \\
\hline \\
AA stacking    & 9.7	 &  \begin{tabular}[c]{@{}c@{}}8.8 (RPA) \cite{Zhou2015} \\ 12.4 (QMC) \cite{Mostaani2015} \end{tabular}             \\
\hline \hline \\
\end{tabular}} \\
\caption{Comparison between LDA energies (using $a=2.461$ \r{A} and $T = 284$ K for the electronic temperature) and previous works. Energies are relative to BG (ABABABAB for the stacking fault, and AB for the saddle point and AA stacking). In configuration AA, the LDA result ($c_d=3.603$ \r{A}) is in between the RPA\cite{Zhou2015} ($a=2.46$ \r{A}, $c_d=3.62$ \r{A}) and QMC\cite{Mostaani2015} ($a=2.460$ \r{A}, $c_d=3.495$ \r{A}) values. RPA and QMC differ by about 40\%, so more theoretical work is needed to determine energy differences more precisely. At the saddle point, LDA ($c_d=3.365$ \r{A}) agrees well with RPA ($c_d=3.42$ \r{A}). The stacking fault value of LDA ($c_d=3.333$ \r{A}) is lower than RPA by 0.07 meV/interface-atom, or about 0.01 meV/atom.}
\label{tab:LDA}
\end{table}

The stacking fault energy is of the order of 0.1 meV/interface-atom. LDA and RPA give similar values, but LDA is lower by 0.07 meV/interface-atom. On the experimental side, Ref.~\citenum{Telling2003} uses anisotropic elasticity theory, revised values for the elastic constants and the experimental data of Refs.~\citenum{Baker1961} and \citenum{Amelinckx1965}, to obtain an average stacking fault energy of 0.14 meV/interface-atom. However, it is an average of 0.05, 0.16, and 0.21 meV/interface-atom, so there is considerable deviation from the mean value. Thus, further experimental work is needed as well.

\subsection{Magnitude of the differences}

The LDA energy difference -0.025 meV/atom at 284 K in Table I corresponds to 0.3 K/atom in temperature units. This suggests that both phases should be observed in the same amount at room temperature, according to the Boltzmann distribution. However, due to the strong intralayer interactions, atoms within one layer move in conjunction. For example, to compare the stability of trilayer ABA vs. ABC, the energy per atom should be multiplied by the number of atoms in the third layer (the whole layer has to move to change the position from A into C, not just one atom). For a flake of 20 nm, this already corresponds to more than $10^4$ atoms, and to energy differences above room temperature. In addition, the exponential character of the Boltzmann distribution translates into large differences in the fraction of each phase (see the Supporting Information, Phase coexistence for small flakes, for further details). This is analogous to the presence of permanent ferromagnetism in cubic Fe, Co and Ni, where the energy of the system depends on the direction of magnetization (property referred to as magnetocrystalline anisotropy) by only $\sim 10^{-3}$ meV/atom (0.01 K/atom)\cite{Halilov1998}. But domains have thousands or millions of atoms, and the energy difference between a domain oriented in different directions can become much larger than room temperature, favoring one direction over another. In the same way, small energy differences per atom in graphene systems can favor either one of the phases (depending on factors like the temperature or the number of layers), as opposed to getting 50\% of each. In making these statements about the stability of BG vs. RG, we are assuming that the difference of the contribution of phonons to the free energy is smaller than the electronic one.

The small energy differences between BG and RG are reflected on the fact that RG has been observed in both natural and artificial samples \cite{Lipson1942,Laves1956}. In graphite crystallized by arcing, Ref.~\onlinecite{Lipson1942} observed 80\% of BG, 14\% of RG and 6\% of random stacking. Shear stress has been known to increase the percentage of RG to 20-30\% \cite{Laves1956,Boehm1964,Freise1963}, and as mentioned earlier, it can lead to long-range rhombohedral order \cite{Nery2020}. Other factors like temperature, pressure and curvature also affect which structure is more stable.
The energy barrier separating one local minimum from another, of about 1.5 meV/atom, is much larger than energy differences between different graphene SSs, which are of the order of 0.01 meV/atom. Thus, different SSs can be similarly stable, and their proportion can depend significantly on the conditions under which a sample is prepared.

\begin{figure}[h!]
\includegraphics[width=1\linewidth]{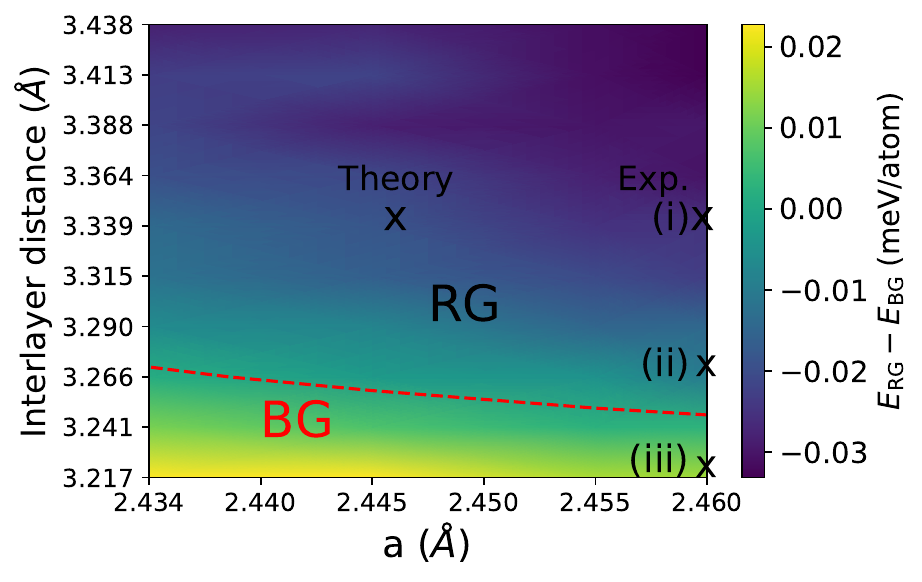}\\
(a)\\
\includegraphics[width=0.9\linewidth]{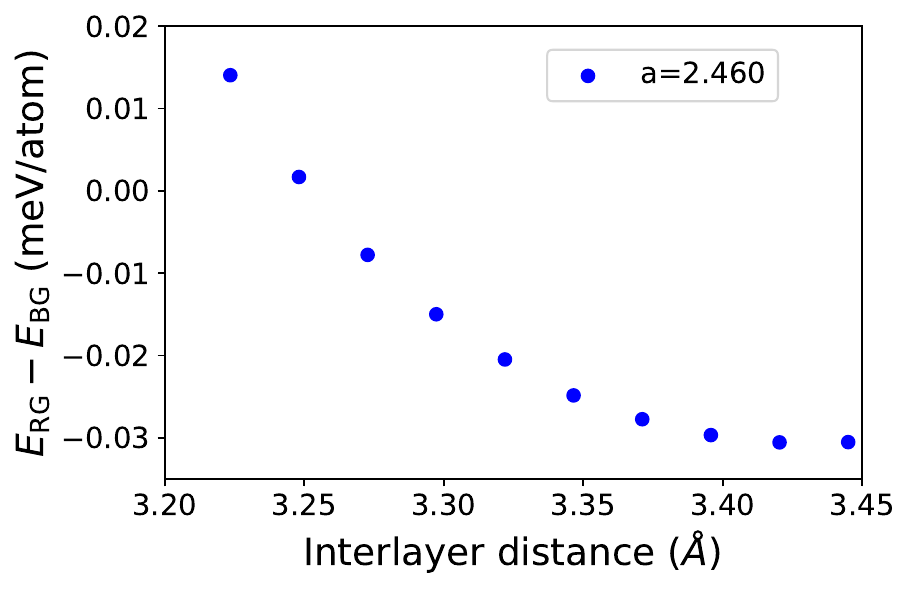}\\
(b)\\
\caption{(a) Bulk energy difference as a function of $c_d$ (interlayer distance) and $a$ at room temperature. At the experimental and theoretical values, RG is more stable. For low enough values of $c_d$, BG has lower energy, so high pressures should favor BG. For 4 layers (see Table~\ref{tab:4L}) and 12 layers (Table~\ref{tab:Ising_fit_roomT}), energies were calculated in (i), (ii) and (iii). (b) Slice of (a) at $a=2.460$ \r{A}.}
\label{fig:c_vs_a}
\end{figure}

\begin{figure}
\centering
	{\includegraphics[width=0.45\textwidth]{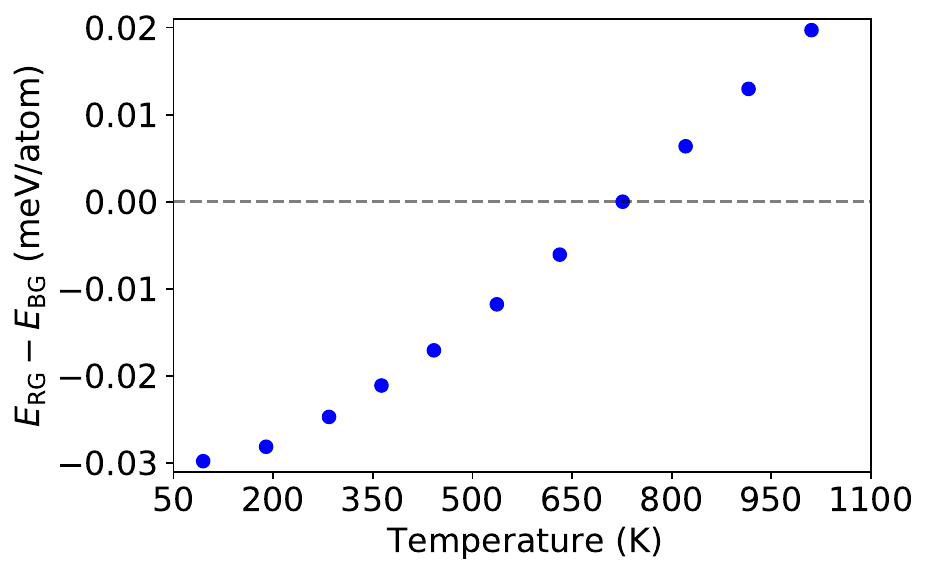}}
\caption{Bulk energy difference as a function of temperature. RG is favored at at low temperatures up to about 730 K. At higher temperatures, BG is favored. The blue dots correspond to the experimental latttice parameters at room temperature \cite{Mounet2005}, while the red dots include experimental thermal expansion values \cite{Entwisle1962}. The results are similar, and thermal expansion will be neglected.}
\label{fig:bulk_vs_T}
\end{figure}

The differences in Table I were calculated using the experimental parameters $a=2.461$  \r{A} (in-plane) and $c_d=3.347$ \r{A} (interlayer distance)\cite{Mounet2005}.  
We also calculated the energy difference at the theoretical values of $a$ and $c_d$ (for which the structure is fully relaxed), and considered the $c_d$ vs. $a$ phase diagram around the theoretical value at room temperature, Fig.~\ref{fig:c_vs_a}. For both the experimental and theoretical values RG is favored, and for most of the range of the parameters (see also Fig.~\ref{fig:c_vs_a_GGA} in the Supporting Information for the GGA-PBE phase diagram, which always favors RG). This figure suggests that thermal expansion may be relevant when considering higher temperatures. However, when using experimental values of $a(T)$ and $c(T)$ [\onlinecite{Entwisle1962}], the energy difference as a function of temperature does not change much (red dots in Fig.~\ref{fig:bulk_vs_T}), and will be neglected in the rest of this work.

Fig.~\ref{fig:c_vs_a} also raises the question of which lattice parameters give energy differences closer to experiments. As indicated earlier, Ref.~\onlinecite{Lipson1942} observed 80\% of BG, 14\% of RG, and 6\% of disordered structures. Samples were produced by arcing (as opposed to exfoliation, which would increase the percentage of RG), indicating that RG is a low energy structure (there are much more disordered configurations, but still account for a lower percentage). Furthermore, Ref.~\onlinecite{Lui2011} uses exfoliation to obtain few layers graphene and looks at the distribution of SSs for $N=4$, for which there are 3 possibilities: ABAB (BG), ABCA (RG) and ABAC (disordered). They observe 85\% of BG, 15\% of RG, and no disordered structure. They argue that these percentages are similar to those of Ref.~\onlinecite{Lipson1942}, and that the complicated pattern of stacking domains cannot be explained by mechanical processing (the stacking of some domains might have still been shifted from BG or disordered to RG though). They conclude that the distribution originates from the pristine structure of graphite. With this in mind, let us look at the $N=4$ energy differences at room temperature (Table~\ref{tab:4L}) with the parameters of (i), (ii), (iii) and (iv) in Fig.~\ref{fig:c_vs_a}(a). Relative to the disorded structure, RG is more stable in (i) (experimental values) and (iv) (theoretical), but is similarly stable or less stable in (ii) and (iii), which suggests the energetics of (i) or (iv) are more representative of experiments. In the rest of the calculations, we will use the experimental values (i).

\begin{table}[h]
{\begin{tabular}{cccc}
\hline \hline
\multicolumn{1}{l}{}  & \multicolumn{3}{c}{Energy difference (meV/atom)} \\
\multicolumn{1}{l}{}  & \begin{tabular}[c]{@{}c@{}} ABAB (BG) \hspace{2mm} \end{tabular} & \begin{tabular}[c]{@{}c@{}} ABCA (RG) \end{tabular} \hspace{2mm}  & ABAC (dis.)\\
\hline
 (i)  & 0      & + 0.007   &  +0.015  \\
\hline \\
(ii)    & 0     & +0.027   &  +0.029      \\
\hline \\
(iii)  & 0	& +0.048 &  +0.043   \\
\hline \\
(iv)  & 0 & +0.013 & +0.018 \\
\hline \hline \\
\end{tabular}} \\
\caption{Energy differences $E - E_{\textrm{BG}}$ for $N=4$, where $E$ can be the energy of ABAB (BG, the reference state), ABCA (RG) or ABAC (disordered). At the lattice parameters (i) and (iv) (see Fig.~\ref{fig:c_vs_a}(a)), RG is favored relative to ABAC, at (ii) they have similar energies, and at (iii) ABAC is favored. In both Refs.~\onlinecite{Lipson1942} and \onlinecite{Lui2011}, RG is favored relative to the disordered structure, suggesting the experimental parameters in (i) or the theoretical ones (iv) better represent the energetics of the SSs. The experimental values will be used in the rest of the calculations.}
\label{tab:4L}
\end{table}

\section{Finite number of layers}

\begin{figure}[h]
\centering
	{\includegraphics[width=0.45\textwidth]{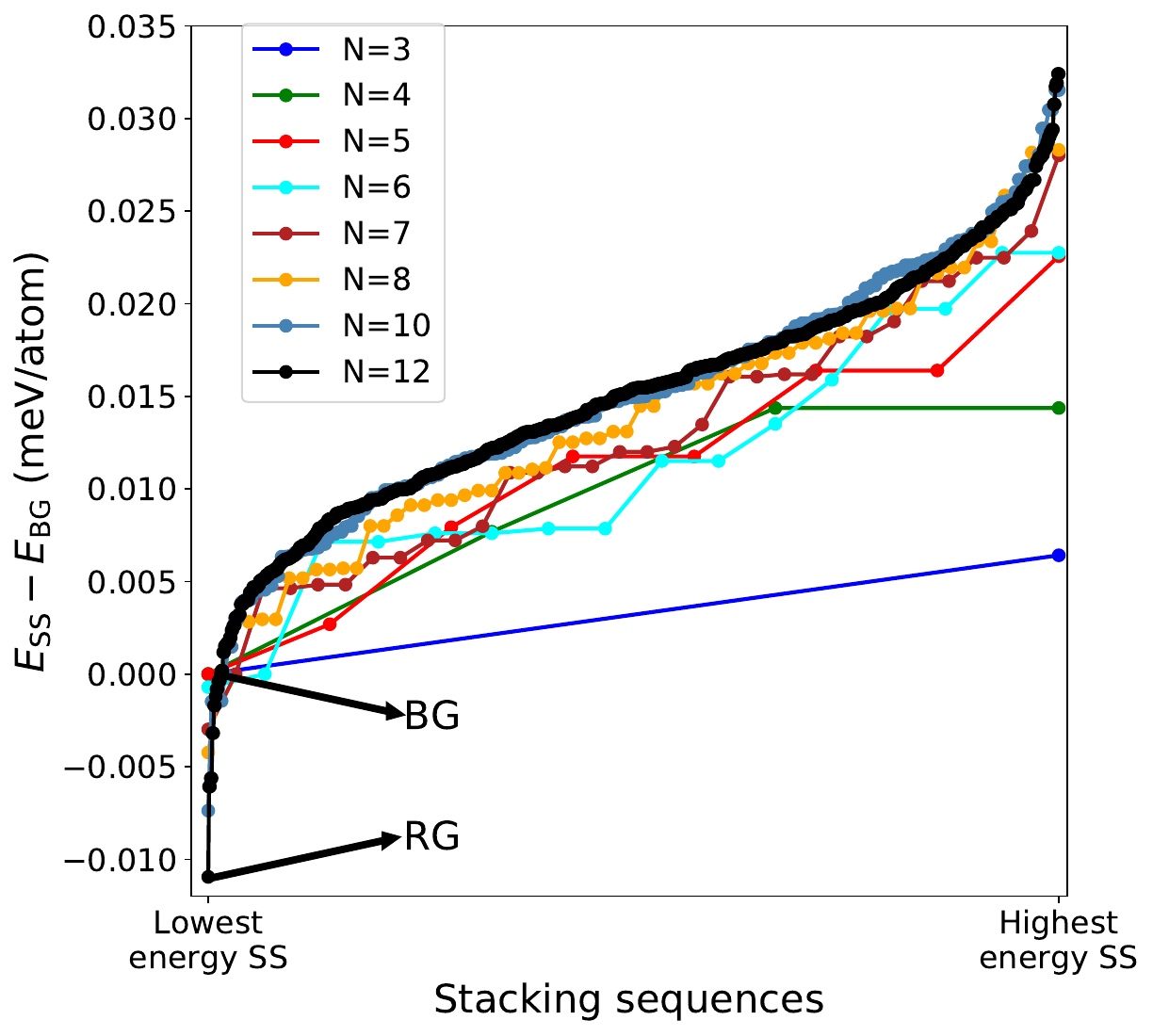}}
\caption{Energy of all stacking sequences at $T = 284$ K for several number of layers $N$, relative to the energy of BG at that $N$ (at each $N$, BG has 0 meV/atom). The lowest energy state is placed at 0 in the $x$ axis, the highest energy state at 1, and all other states at placed equidistantly in between. There are $2^{N-2}$ states for each $N$. For $N=3,4,5$, BG is the most stable structure, while for $N \geq 6$ it is RG.
In addition, the low energy structures are mixed BG-RG systems with soft interfaces. See Fig.~\ref{fig:configs_T284} for more details. The energy distribution can be fit with an Ising spin model (see Fig.~\ref{fig:fit_coeffs}).}
\label{fig:energy_distribution_all}
\end{figure}

In order to better understand the energy distribution of SSs, let us start by looking at $T=284 K$ for different amount of layers $N$ (Fig.~\ref{fig:energy_distribution_all}). For each $N$, the reference state at 0 meV/atom energy is the corresponding AB-stacked sequence. The SS with the lowest energy is placed in the $x$ axis at 0, the highest energy sequence is placed at 1, and the other configurations are placed equidistantly in between. Degenerate structures are included (like ABAC and ABCB, due to up and down symmetry), to correctly capture the energy distribution. For example, for $N=3$ there are only two possibilities, ABA (BG) and ABC (RG), and BG has lower energy. Up to $N=5$, BG has the lowest energy. At $N=6$, there is a configuration with energy slightly lower than 0, which is RG. It is interesting to note that the lowest energy corresponds to BG or RG for all $N$, and that the energy distribution seems to converge to a universal curve.

\subsection{Spin notation}

A given SS can be mapped to a spin chain, putting a variable $\s_i$ with value +1 or -1 at each interface: for a layer with a given position (A, B or C), if the upper layer's position is the next one in the periodic sequence ABC, then $\sigma=+1$, and -1 otherwise. See Table~\ref{tab:spin_notation} for some examples. This makes visualization easier, and will be used next for low energy structures, but its main purpose is to later fit the DFT energies with an Ising model.

\begin{table}
{\begin{tabular}{ccc}
\hline \hline
 & \begin{tabular}[c]{@{}c@{}} Standard ABC \\ notation \end{tabular} &  \begin{tabular}[c]{@{}c@{}} Spin $+-$ \\ notation \end{tabular} \\
\hline
BG & {\footnotesize ABABABAB} & {\footnotesize $+-+-+-+$} \\
\hline \\
RG & {\footnotesize ABCABCABC} & {\footnotesize $++++++++$} \\ 
\hline \\
\begin{tabular}[c]{@{}c@{}} Soft BG-RG \\ interface \end{tabular} &  {\footnotesize ABAB[\textbf{AB}]CAB}  & {\footnotesize $+-+-++++$}  \\
\hline \\
\begin{tabular}[c]{@{}c@{}} Hard BG-RG \\ interface \end{tabular}  & {\footnotesize BABAB][CBAC}  & {\footnotesize $-+-++---$}  \\
\hline \hline \\
\end{tabular}} \\
\caption{Illustration of the spin notation for several SSs.}
\label{tab:spin_notation}
\end{table}

\subsection{Soft and hard interfaces}
 
\begin{figure}[h]
\centering
	{\includegraphics[width=0.45\textwidth]{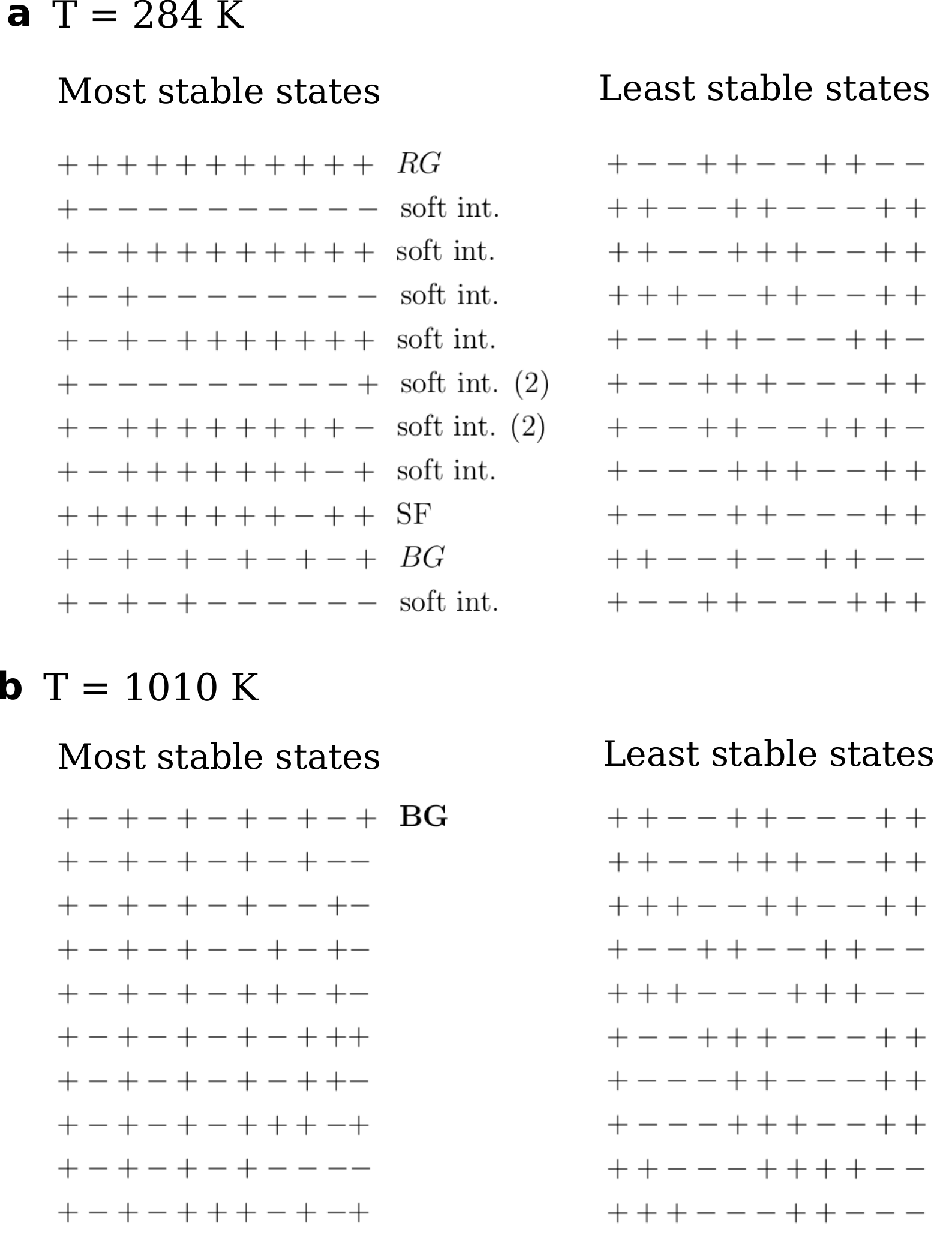}}
\caption{\textbf{a}, At $T = 284$ K, the low energy states are RG, BG and mixed BG-RG states with soft interfaces. The number between brackets indicates the number of soft interfaces. SF stands for stacking fault. \textbf{b}, At $T = 1010$ K, the low energy states have mostly alternating $+,-$ signs. The high energy states are similar in both cases, with 2 or 3 consecutive $+$ signs followed by 2 or 3 $-$ signs (hard RG interfaces).}
\label{fig:configs_T284}
\end{figure}

To characterize the low energy states, let us notice that there are two types of interfaces between BG and RG, that we refer to as soft and hard. A SS with a soft interface has the form ABAB[\textbf{AB}]CABC. Starting from the left, the structure is BG (up to the bracket ]), while starting from the right the structure is RG (up to the bracket [). Thus, the layers in between [ and ] are both part of the BG and RG sequences. The form of a hard interface is ABABAB][CBACBA, so no layers are shared. The energy of the soft interface is 0.07 meV/atom, while that of the hard interface is 0.18 meV/atom, which is higher, as expected (see Table~\ref{tab:energies_int}, which also shows the energy of other interfaces). The most and least stable states for $N=12$, at $T=284$ K and $T=1010$ K, can be observed in Fig.~\ref{fig:configs_T284}. At room temperature, the low energy states after RG have soft interfaces. The form $+----$ is also a soft interface, since it corresponds to A[BA]CB. The only exception is $++++++++-++$, which would correspond to a stacking fault. The least favored states have several consecutive $+$ signs, followed by consecutive $-$ signs, which correspond to hard RG interfaces.The form $+----$ is also a soft interface, since it corresponds to A[BA]CB. The only exception is $++++++++-++$, which would correspond to a stacking fault. Therefore, the most stable structures are BG and RG (together with mixed BG-RG structures with soft interfaces), while random structures are disfavored, in accordance with experiments\cite{Lipson1942}. This is one of the main results of our work.

\begin{figure}[h]
\centering
	{\includegraphics[width=0.45\textwidth]{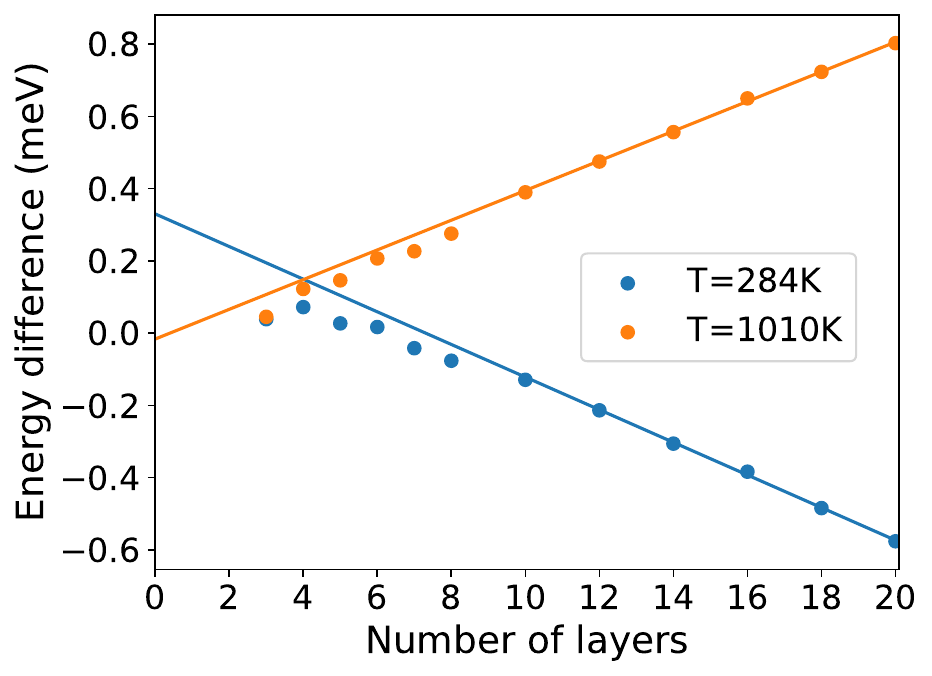}}
\caption{Energy difference between RG and BG for different number of layers at $T = 284$ K (blue) and $T = 1010$ K (orange). The straight lines are a fit for $N \geq 12$, and their slope give the energy difference for bulk. At room temperature the difference is -0.023 meV/atom, favoring RG, while at $T = 1010$ K it is +0.021 meV/atom, favoring BG (which coincide, within error, with the values of Table~\ref{tab:functionals}).} 
\label{fig:diff_vs_layers}
\end{figure}

As the number of layers increases, the energy of RG decreases relative to BG. In the bulk limit ($N \rightarrow \infty$), the difference is -0.023 meV/bulk-atom, favoring RG. This value corresponds to the slope of the blue line ($T=284$ K) in Fig.~\ref{fig:diff_vs_layers}, which shows the difference between BG and RG for different $N$ at 284 $K$ and 1010 K. At 1010 K, BG is more stable (for any number of layers), which is consistent with experiments. For examples, graphite samples where X-ray diffraction showed presence of both BG and RG had to be heated up above 1300 $^\circ$C [\onlinecite{Laves1956}] or 1400 $^\circ$C [\onlinecite{Boehm1964}] to start transforming to BG. They changed completely to BG at about 2700-3000 $^\circ$C [\onlinecite{Boehm1964}]. It is interesting to note that the difference in surface energy (the $y$-intercept) is about 0.08 meV/surface-atom (RG has larger surface energy than BG) at $T = 284$ K, but it essentially goes to 0 (-0.004 meV/surface-atom) at $T = 1010$ K.

\subsection{Phase diagram}

\begin{figure}[h]
\centering
	{\includegraphics[width=0.45\textwidth]{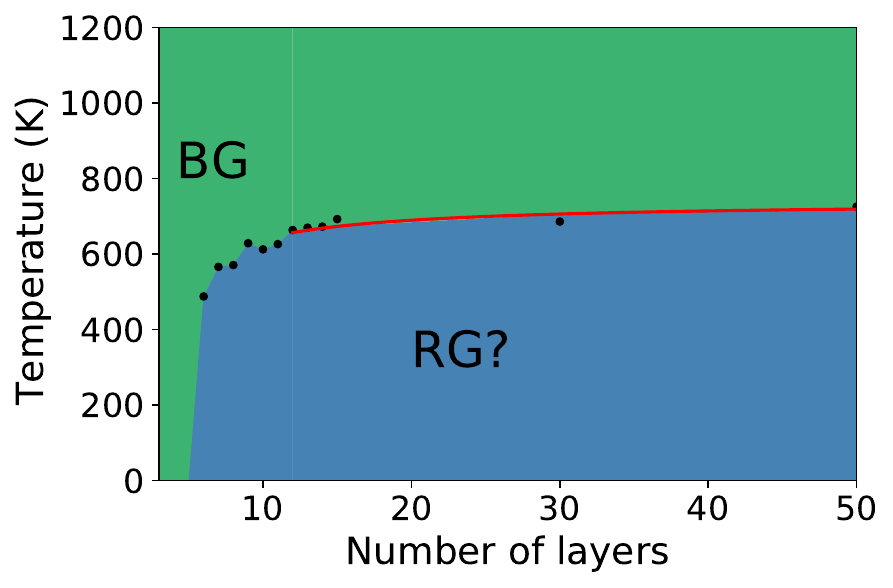}}
\caption{Temperature vs. Number-of-layers phase diagram of the most stable SS. At high temperatures, BG is the most stable structure, and also at room temperature for a small amount of layers. At room temperature for $N \geq 6$, RG is the most stable structure, which seems inconsistent with experiments. However, since BG is the most stable structure for small $N$, nucleation could lead to BG for larger $N$. In addition, natural graphite is formed at higher temperatures, at which BG is more stable, and could remain locked at a local minimum at room temperature. The red curve corresponds to a $1/N$ fit for $N \geq 12$ (see the Supporting Information for details).}
\label{fig:phase_diagram}
\end{figure}

Since bulk RG is more stable at room temperature, and bulk BG at higher temperatures, a phase transition occurs at some temperature in between. More in general, Fig.~\ref{fig:phase_diagram} shows a temperature vs $N$ phase diagram. Our results are consistent with the fact that growth of few layers graphene typically leads to BG. But they seem at odds with BG being the most common structure at room temperature. However, graphite likely forms in the 350-700 $^\circ$C range (650-1000 K), which favors BG (Table I, Fig.~\ref{fig:bulk_vs_T}), and under high pressures\cite{Nover2005}, which also favors BG according to our calculations (Fig.~\ref{fig:c_vs_a}). At higher temperatures, BG becomes more stable (RG transforms to BG,  as we mentioned in the previous paragraph), and so it is possible that it remains in this local minimum as it cools down (even if RG is the most stable structure at room temperature). In addition, if BG is the most stable SS for a few layers, due to nucleation, further layers will continue stacking in the AB order. To be more precise, if the starting sequence consists of a few layers of AB stacked graphene $+-...+-$, then $+-...+-+$ has less energy than $+-...+--$. 
Therefore, it is not unreasonable that beyond certain amount of layers, RG becomes the most stable phase.  As mentioned earlier, in our calculations BG is favored up to $N=5$, and for $N=4$, Ref.~\onlinecite{Lui2011} also obtained that BG is favored. In terms of the stability for $N \geq 6$, Ref.~\onlinecite{Yang2019} has observed up to 27 layers of RG. But the final structure depended strongly on the direction of transfer (due to the induced shear stress), so it is hard to draw definite conclusions about the stability of BG vs. RG under no external factors.

\subsection{Ising fit}

To further understand the energy distribution of SSs, using the spin notation for a system of $N$ layers ($N-1$ spins), we can consider an Ising type of Hamiltonian,

\be
\begin{split}
H=A_0(N-1) & + A_1\sum_{i=1}^{N-2} \s_{i} \s_{i+1} + A_2\sum_{i=1}^{N-3} \s_{i} \s_{i+2} \\
& + A_3\sum_{i=1}^{N-4} \s_{i} \s_{i+3} + \dots,
\end{split}
\label{eq:Ising}
\ee

\ni $A_i$ corresponds to the $i+1$ nearest neighbor (n.n.) interaction energy. In particular, $A_0$ is just the n.n.  (1st) interaction energy and is independent of the SS.

\begin{figure}
\centering
	{\includegraphics[width=0.45\textwidth]{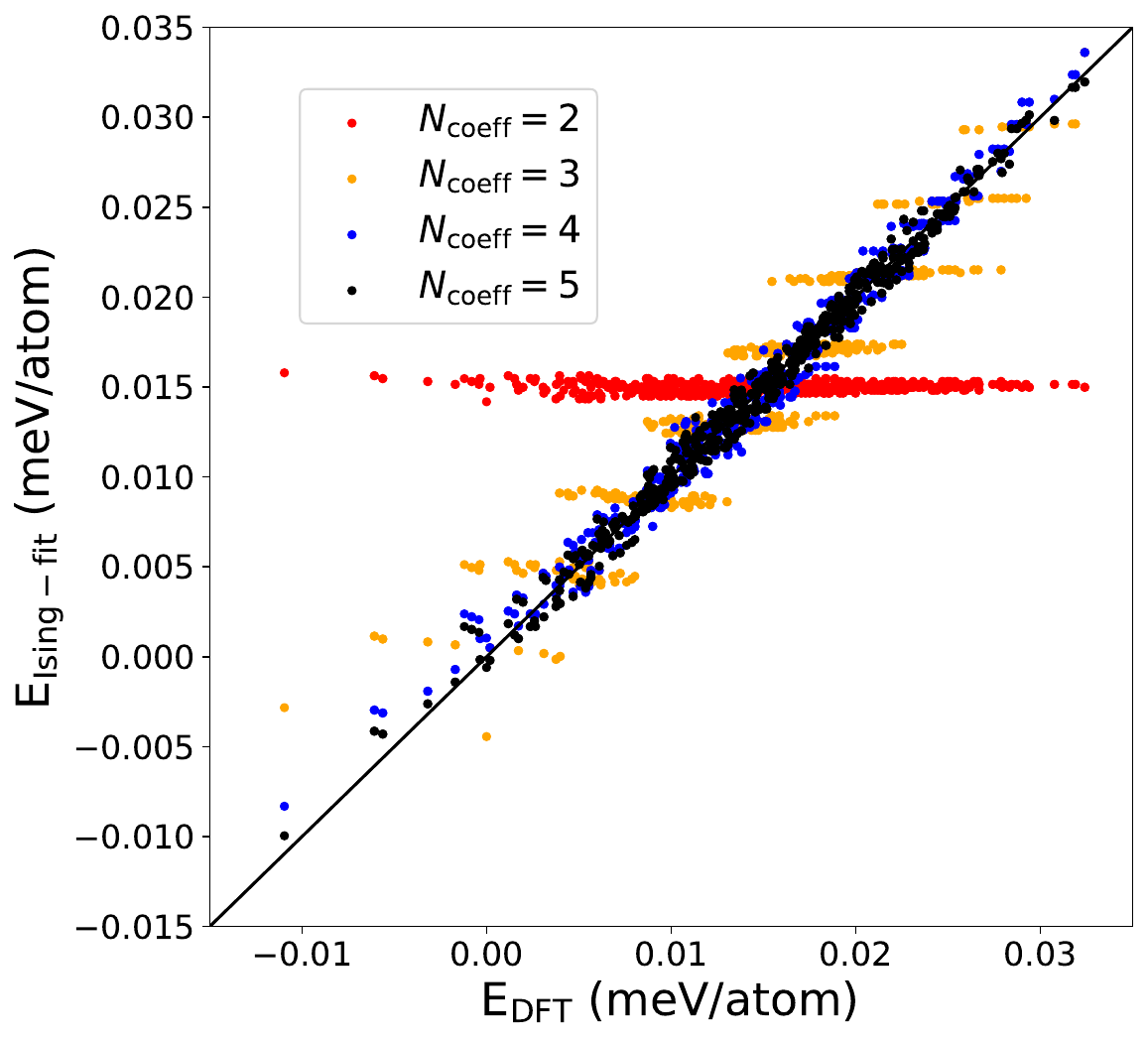}}
\caption{Fit of the DFT energies using the Ising model, Eq.~\eqref{eq:Ising}, for $N=12$ layers at $T = 284$ K. RG is the lowest energy state, and by definition, $E_\textrm{BG}=0$. To favor both BG and RG, $|A_2| \gg |A_1|$ and $A_2<0$, so $N_\textrm{coeff}=2$ does not work well. $N_\textrm{coeff}=4$ works significantly better than $N_\textrm{coeff}=3$, and is similar to $N_\textrm{coeff}=5$. $A_3$ is also negative, so sequences with several consecutive $+$ or $-$ (with rhombohedral order) are favored.}
\label{fig:fit_coeffs}
\end{figure}

For large enough $N$, when the interaction between the surfaces becomes small, $A_i$ should become constant, since an extra layer just adds an additional $A_i$ term. 
 This model is used to fit the DFT energies, as shown in Fig.~\ref{fig:fit_coeffs} for $N=12$ (the results at $T=1010 K$ can be seen in Fig.~\ref{fig:fit_coeffs_T1010}). For both BG and RG to be favored over other structures, $|A_2| \gg |A_1|$ and $A_2 < 0$, so using only $A_0$ and $A_1$ (red dots) gives a very bad fit. Using coefficients up to $A_3$, the fit is good, and including $A_4$ does not improve it much. To keep the model simpler and shorter ranged, we use $A_i,0 \leq i \leq 3$. 
In the Supporting Information (Figs.~\ref{fig:N=20}, \ref{fig:fit_smaller_N}, and \ref{fig:coeffs_vs_N}), we show that the $N=12$ fit also works well for lower and larger $N$. That is, it mantains the relative order of the DFT energies of the SSs. 

\begin{table}
{\begin{tabular}{ccc}
\hline \hline
 & T = 284 K &  T = 1010 K \\
\hline
$A_0$ & 0.0327 & 0.0423 \\
\hline \\
$A_1$ & 0.0019 & 0.0257 \\ 
\hline \\
$A_2$ &  -0.0497 & -0.0288  \\
\hline \\
$A_3$  & -0.0164  & -0.0062  \\
\hline \hline \\
\end{tabular}} \\
\caption{Ising fit coefficients for $N=12$, at room temperature and 1010 K. At low temperatures, $|A_2| \gg |A_1|$ and both BG and RG have low energies. Also, $A_2$ is about three times larger than $A_3$, and the low energy states have soft-interfaces. At 1010 K, $|A_1|$ is larger, which disfavors RG.}
\label{tab:Ising_fit}
\end{table}

\section{Concluding remarks}

We have shown that for bulk graphite, at the experimental values of the lattice parameter $a$ and interlayer distance $c_d$, RG is more stable than BG for different functionals with electrons at room temperature, while at 1000 K and higher temperatures BG is more stable. Although the energy differences are very small, they are similar for all functionals (and in particular, have the same sign), suggesting they are meaningful and not just arbitrary fluctuations. We then argue, as it occurs in magnetic systems, that the energy difference per atom has to be multiplied by the number of atoms in the sample, resulting in a phase preference at a given temperature (as opposed to 50\% of each phase, or even random stacking sequences (SSs) as well). For LDA, we also showed that RG is more stable at the theoretical values of $a$ and $c_d$. A phase diagram of $c_d$ vs. $a$ shows that this holds around these values as well, unless for example the interlayer distance is contracted about 3\%. This suggests low pressures are better to grow RG. In order to determine the most likely value of $c_d$, we looked at the energetics of $N=4$ and compared to experimental data, which showed no disordered structure. While the experimental $c_d$ favored RG over the disordered SS, lower values of $c_d$ make them equally likely or even disfavors RG. In addition, the distribution of graphite obtained through arcing shows that RG should have lower energy than disordered structures.

We also looked at the stability for a finite amount of layers $N$, and obtained a temperature vs. $N$ phase diagram. At temperatures of about 700 K or higher, BG is more stable for any $N$. This is consistent with experiments, since RG disappears when heating up samples with both BG and RG at high enough temperatures. At room temperature, BG is more stable for a few layers, and then RG becomes more stable. Although this seems at odds with BG being the most abundant natural phase, graphite is formed at higher temperatures and pressures, so it could get locked at a local minimum when cooling down. Furthermore, since BG is favored for a few layers, and the interface energy is significant, nucleation could lead to further layers of BG.

We then calculated the energy for all SSs for $N=12$, and noticed that the low energy states correspond to RG, BG and mixed BG-RG systems with soft interfaces (using the experimental parameters). 
Finally, we showed that the energy distribution can be fit with an Ising model, and that the $N=12$ fit can be used to determine the stacking order for larger $N$ (without need to calculate the energy of thousands or millions of SSs). 

Determining the very small energy differences in multilayer graphene systems is still a challenging problem, and more experimental and theoretical studies are needed. Our work helps to better understand their stability at different electronic temperatures and for different number of layers, and serves as a starting point to add external factors like curvature, doping or an electric field in first-principles calculations.

\section*{Acknowledgments}

We acknowledge support from the European Union’s Horizon 2020 research and innovation programme Graphene Flagship under grant agreement No 881603. M.C. also acknowledges support from
Agence nationale de la recherche (Grant No. ANR-19-CE24-0028). The authors would like to thank Stony Brook Research Computing and Cyberinfrastructure, and the Institute for Advanced Computational Science at Stony Brook University for access to the high-performance SeaWulf computing system, which was made
possible by a \$1.4M National Science Foundation grant (\#1531492).

\clearpage

\appendix
\renewcommand\thefigure{S\arabic{figure}} 
\setcounter{figure}{0} 

\renewcommand\thetable{S\arabic{table}} 
\setcounter{table}{0} 

\renewcommand\theequation{S\arabic{equation}} 
\setcounter{equation}{0}

\section*{Supporting Information}

\ni \textbf{DFT calculations.} Calculations were carried out with Quantum Espresso (QE)\cite{QE2017}, using an LDA functional unless otherwise specified. $k$-grids for vacuum calculations were done with $80 \times 80 \times 1$ $k$-grids. For the planewaves, a cutoff of 70 Ry was used in Tables\ref{tab:functionals} and \ref{tab:4L}, and in Fig.~\ref{fig:c_vs_a}. In the other cases, 60 Ry were used. The cutoff for the charge-density was eight times these values. It is worth pointing out that to converge energies themselves (as opposed to their differences) at the 0.001 meV/atom level, much higher energy cutoffs are needed.


When referring to an interface or surface, energy differences corresponds to an area, and are divided by 2 (since there are 2 atoms in the primitive cell), and we write them in meV/interface-atom units. Another common unit in the literature is mJ/m$^2$. Otherwise, the energy differences are divided by the total amount of atoms in the structures that are being compared, and the units are meV/atom.\\

\begin{figure}[h]
\centering
	{\includegraphics[width=0.45\textwidth]{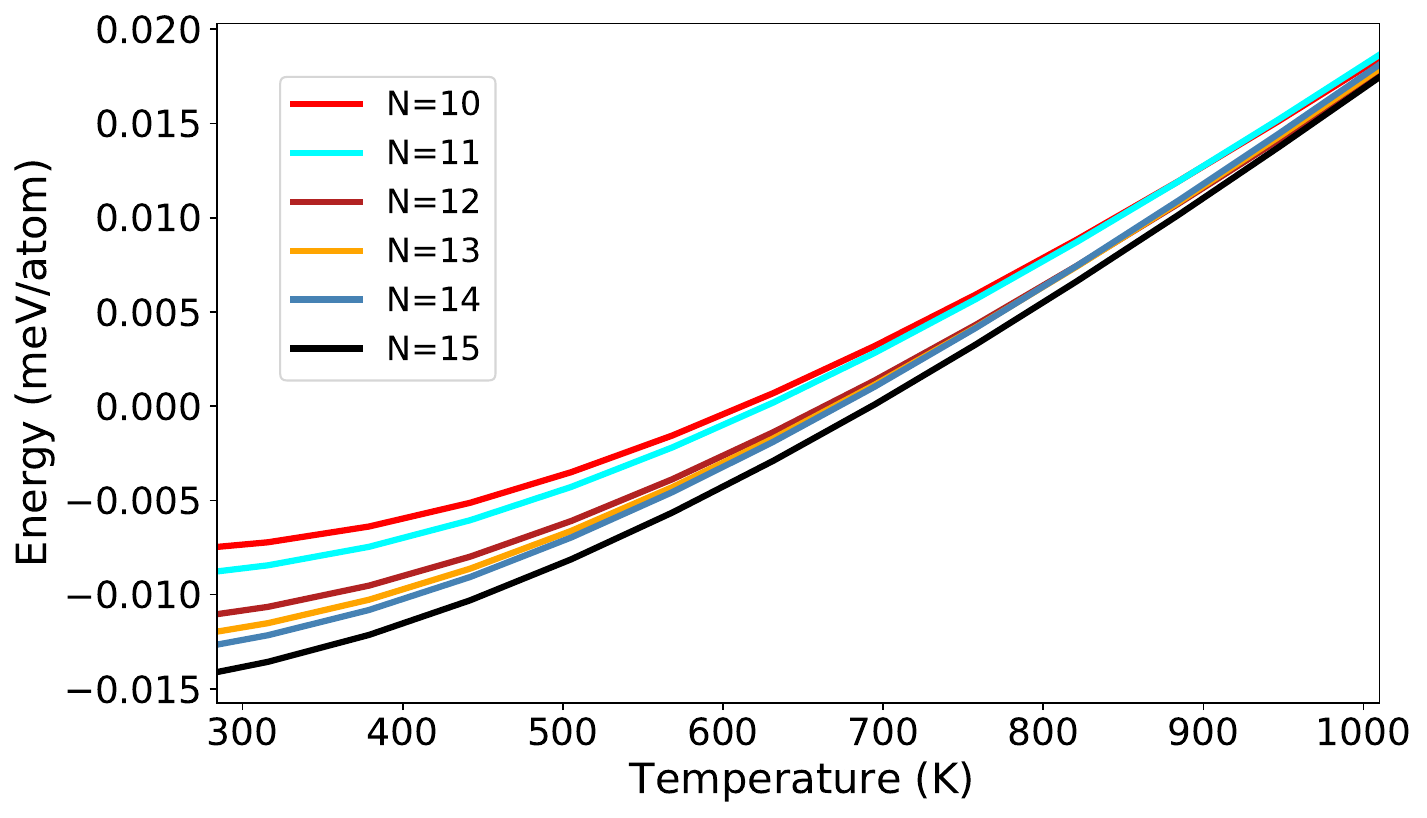}}
\caption{Energy difference between RG and BG per atom as a function of temperature $T$ for several $N$. As a rough estimate, the energy increases linearly with $T$. This is used to obtain the $1/N$ model, for $N \geq 12$, for the critical temperature of Fig.~\ref{fig:phase_diagram} (see Eq.~\eqref{eq:1/N}). 
}
\label{fig:energy_vs_T}
\end{figure}

\ni \textbf{Transition temperature $\mathbf{1/N}$ model.} In Fig.~\ref{fig:phase_diagram} we used a $1/N$ model to fit the transition temperature $T_c$ from RG to BG. For a fixed temperature, the energy difference $\Delta E = E_\textrm{RG}-E_\textrm{BG}$ increases linearly beyond certain amount of layers (10 or 12 from Fig.~\ref{fig:diff_vs_layers}), when there is no more interaction between the surfaces. Let us assume, to simplify, that the difference $\Delta E = E_\textrm{RG}-E_\textrm{BG}$ is also linear with temperature (see Fig.~\ref{fig:energy_vs_T}). From these two conditions, we can write

\be
\Delta E(N,T)=B_{11} + B_{12} N + B_{21} T + B_{22} N T.
\ee

\ni $T_c(N)$ is given by $\Delta E(N,T_c)=0$. For large enough $N$,

\be
T_c = \f{-B_{11}}{N} + B_{12},
\label{eq:1/N}
\ee

\ni as we wanted to show.\\

\ni \textbf{Phase coexistence for small flakes}: According to the Boltzmann distribution, the fraction $p_i$ of a given SS is

\be
\begin{split}
p_i & =\f{e^{-\Delta E_\textrm{eff,i}/T}}{Z}\\
\textrm{with} \hspace{2mm} Z & =\sum_j^{2^{N-2}} e^{-\Delta E_\textrm{eff,j}/T},
\end{split}
\ee

\begin{figure}[h]
\centering
	{\includegraphics[width=0.45\textwidth]{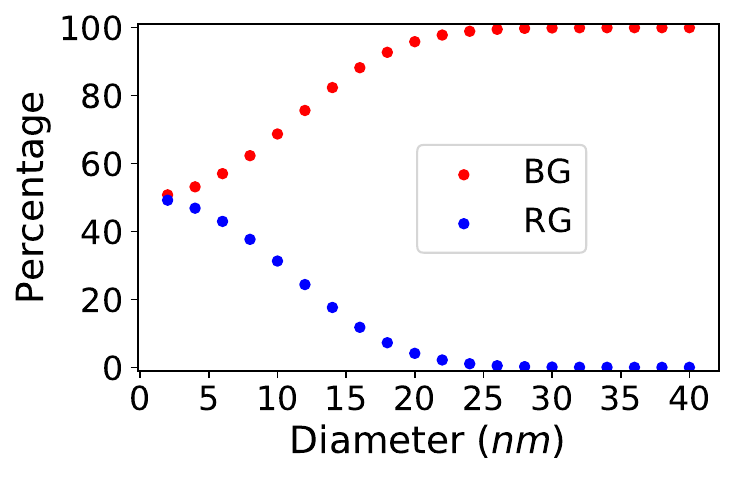}}
\caption{Percentage of trilayer BG and RG, as a function of the diameter of the third layer. Although the energy difference per atom is small, 0.006 meV/atom (0.074 K/atom), small diameters already correspond to thousands of atoms, which lead to an effective energy difference that is (in temperature units) larger than room temperature.}
\label{fig:percentage_3L}
\end{figure}

\ni where $\Delta E_\mathrm{eff,i}$ is the effective energy of configuration $i$ relative to BG. For example, in the case of trilayer graphene, $\Delta E_\mathrm{eff,i} \sim N_a\Delta E_\mathrm{i}$, where $N_a$ is the amount of atoms in the third layer and $\Delta E_\mathrm{i}= E_i - E_\textrm{BG}$ is the energy difference per atom with respect to BG. Fig.~\ref{fig:percentage_3L} shows the percentage of ABA and ABC at room temperature as a function of the diameter of the third layer. For diameters below 20-25 nm, both BG and RG should be observed, while for larger diameters only BG.\\

\ni \textbf{Measurement of $E_\textrm{RG}-E_\textrm{BG}$}: Based on the previous discussion, a possible experiment to determine $E_\textrm{RG} - E_\textrm{BG}$ could consist of using two large layers of AB graphene, and a third smaller layer. The system would be heated up so that the third layer can translate on the plane, and then cooled down so that it relaxes back to ABA or ABC. Doing this repeatedly, one could determine the relative frequency of each phase, and thus the energy difference. Flakes should not be smaller than a few nm though, since then the assumption of infinite length is not valid anymore \cite{Belenkov2001}. A drawback of this method is that the sample might have to be heated significantly, and the energy difference would correspond to the higher temperature that activates translations, as opposed to room temperature.\\


\ni \textbf{Soft and hard interfaces.} To estimate the soft and hard interfaces, we use the $N=12$ energies. The interface energy (soft or hard) between BG and RG can be written as

\be
E_\textrm{int}=[E_\textrm{BG-RG} - E_\textrm{BG}/2 - E_\textrm{RG}/2]/2,
\ee

\ni where $E_\textrm{BG-RG}$ is the energy of the system with 6 BG layers and 6 RG layers (in the soft case, the limit ``$|$" is put in the middle of the shared layers: 01010[1$|$0]21021).  $E_\textrm{BG}$ and $E_\textrm{RG}$ are divided by 2 because there are 6 layers of each. In doing so, only one BG and one RG surface are is subtracted, just as in the BG-RG system. Thus, the reminding energy corresponds to the interface energy. The external division by 2 gives the interface energy per atom. For soft and hard interfaces in BG-BG or RG-RG systems, we calculate

\be
E_\textrm{int}=[E_\textrm{SS-SS} - E_\textrm{SS}]/2
\ee

\ni where SS=BG or SS=RG. Results are included in Table~\ref{tab:energies_int}.\\

\ni \textbf{Ising model: further analysis.} The Ising model can be used to gain more insight into which are the low energy states in terms of its coefficients (see Table~\ref{tab:Ising_fit_roomT}).

$A_1>0$ favors BG and states that are basically BG, but with two consecutive equal signs in some part of the sequence (e.g., $+-+-+--+-+$). It disfavors RG and RG-RG states with hard interfaces.

$A_2<0$ favors both BG and RG as mentioned earlier, and interestingly, also mixed BG-RG with soft interfaces.

$A_3<0$ favors states with one sign followed by two of the other sign (e.g., $+-++-++-$) and also RG, while it disfavors BG and hard RG-RG interfaces.

As mentioned earlier, at room temperature $|A_2| \gg |A_1|$ (see Table~\ref{tab:Ising_fit}). It was implied that $|A_3|$, which corresponds to more distant neighbors (4th n.n), is also small compared to $|A_2|$. At 1010 K, $|A_1| \sim |A_2|$, and since $A_1>0$ disfavors RG while $A_2<0$ favors it, RG lies in the middle of the spectrum (slightly below actually, since $A_3<0$, but it is small relative to $A_1$ and $A_2$).

We also calculated the energy for all SSs at room temperature for the parameters (ii) and (iii) in Fig.~\ref{fig:c_vs_a} (see Table~\ref{tab:Ising_fit_roomT}). First of all, the fact that $A_0$ increases for lower $c_d$ shows that BG is becoming more stable relative to the other states. At (ii), $A_2$ (in absolute value) is larger than $A_1$ and $A_3$, but by factor of about 2 and 3 only. So there are soft interfaces for low energy states, but aside from RG, the very lowest states are the ones favored by $A_1$ (e.g. $+-+-+--+-+$). At (iii), $|A_1/A_2|$ increases and the soft BG-RG states go down a little bit lower. This also occurs when increasing the temperature. On the other hand, in Table~\ref{tab:Ising_fit} $|A_3/A_2|$ becomes smaller, while here it remains essentially the same. In any case, as pointed in the main text, Refs.~\onlinecite{Lipson1942} and \onlinecite{Lui2011} suggest that the LDA energies corresponding to the experimental parameters (i) are closer to the actual experimental energies.

\begin{figure}[h]
\centering
	{\includegraphics[width=0.45\textwidth]{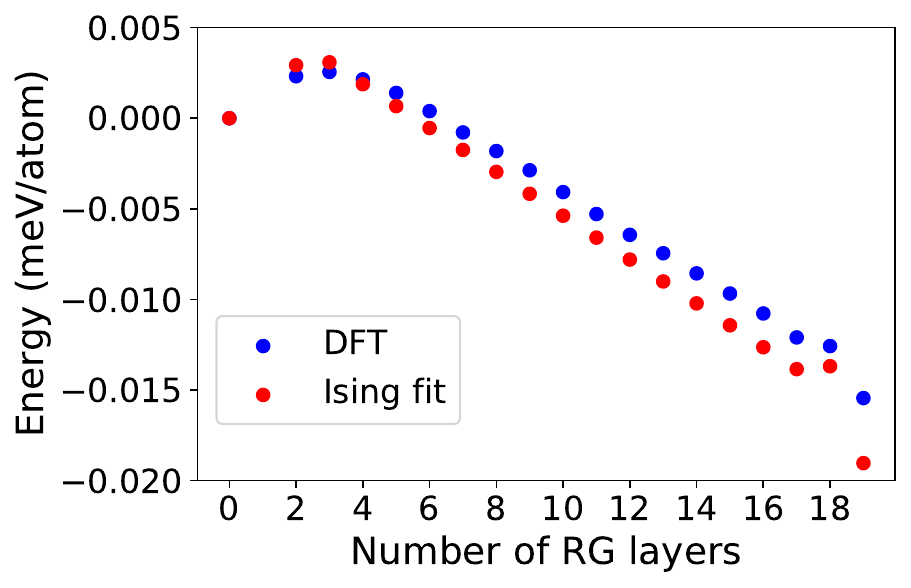}}
\caption{Energy difference in a BG-RG system of $N=20$ layers with a soft-interface, as a function of the number of RG layers. To be more precise, configurations are (0) $+-...-+$, (2) $++-+...+-$, (3) $+++-+...-+$,...,(18) $++...++-$, (19) $++...++$, so it corresponds to the number $N_+ \neq 1$ of consecutive $+$ signs ($N_+=1$ just corresponds to BG). The blue dots correspond to the direct DFT result, while the red dots are obtained using the $N=12$ Ising fit. The fit is shifted, setting the energy of BG at 0. It adequately determines the relative energies, suggesting that it can be indeed used to determine the order of SSs for $N>12$. In addition, we see that RG is the most stable structure for large $N$. But BG is more stable for a few layers. Is a SS with many layers of RG and a few layers of BG the most stable structure? The answer is no, because there is a high amount of energy in the BG-RG interface, of about 0.07 meV/interface-atom. The energy difference between RG and BG is 0.006 meV/atom for $N=3$ (and smaller for $N=4,5$). When substituting 3 layers of RG with BG, the energy roughly decreases 0.006 $\times$ 3 meV, but increases 0.07 meV, resulting in a more energetic state.}
\label{fig:N=20}
\end{figure}

\begin{figure*}[h]
\centering
	{\includegraphics[width=0.8\textwidth]{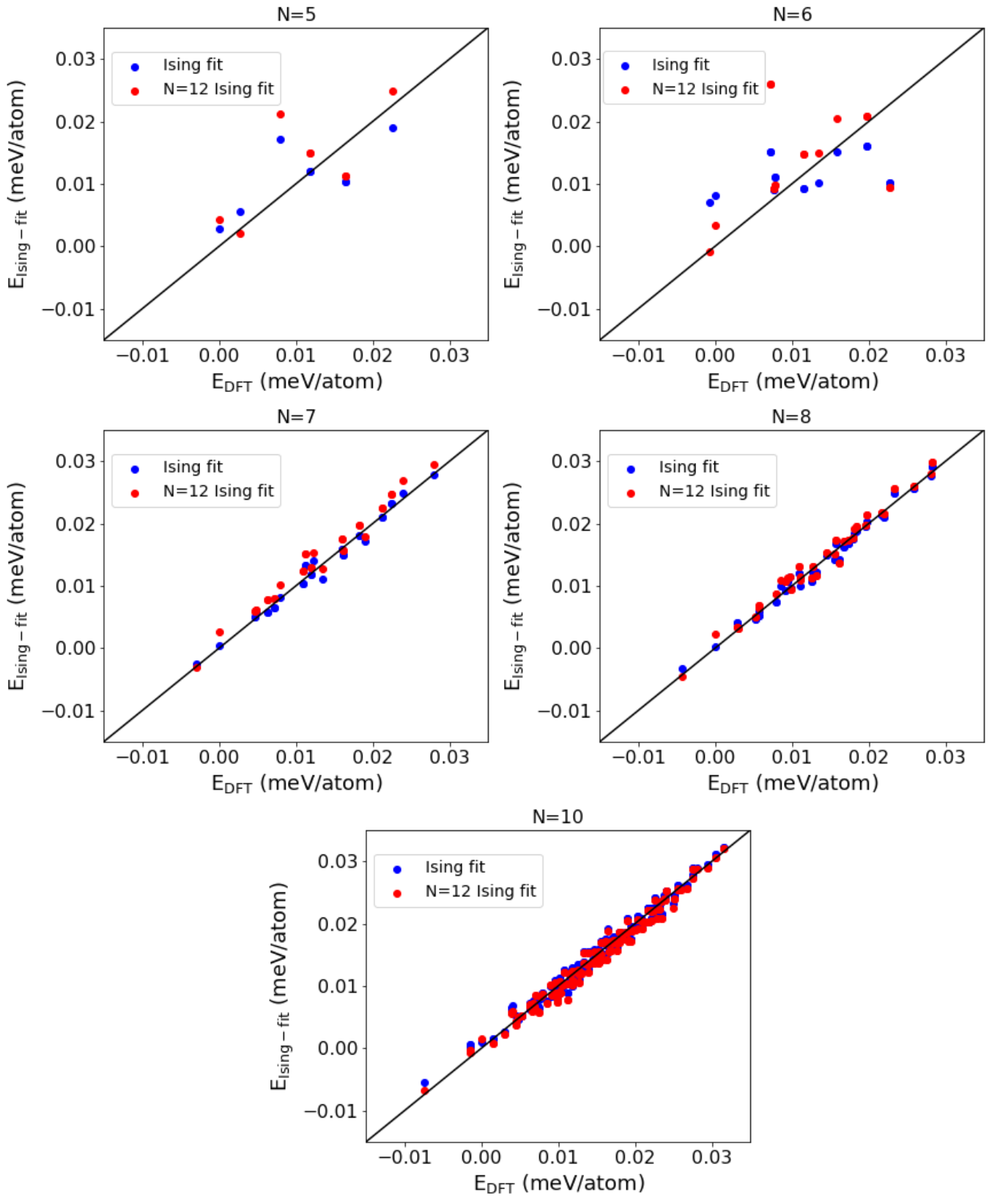}}
\caption{Comparison between the Ising fit for several $N$, with the coefficients resulting from the $N=12$ fit. The similarity shows the fit is transferable.}
\label{fig:fit_smaller_N}
\end{figure*}

\begin{figure*}[h]
\centering
	{\includegraphics[width=0.9\textwidth]{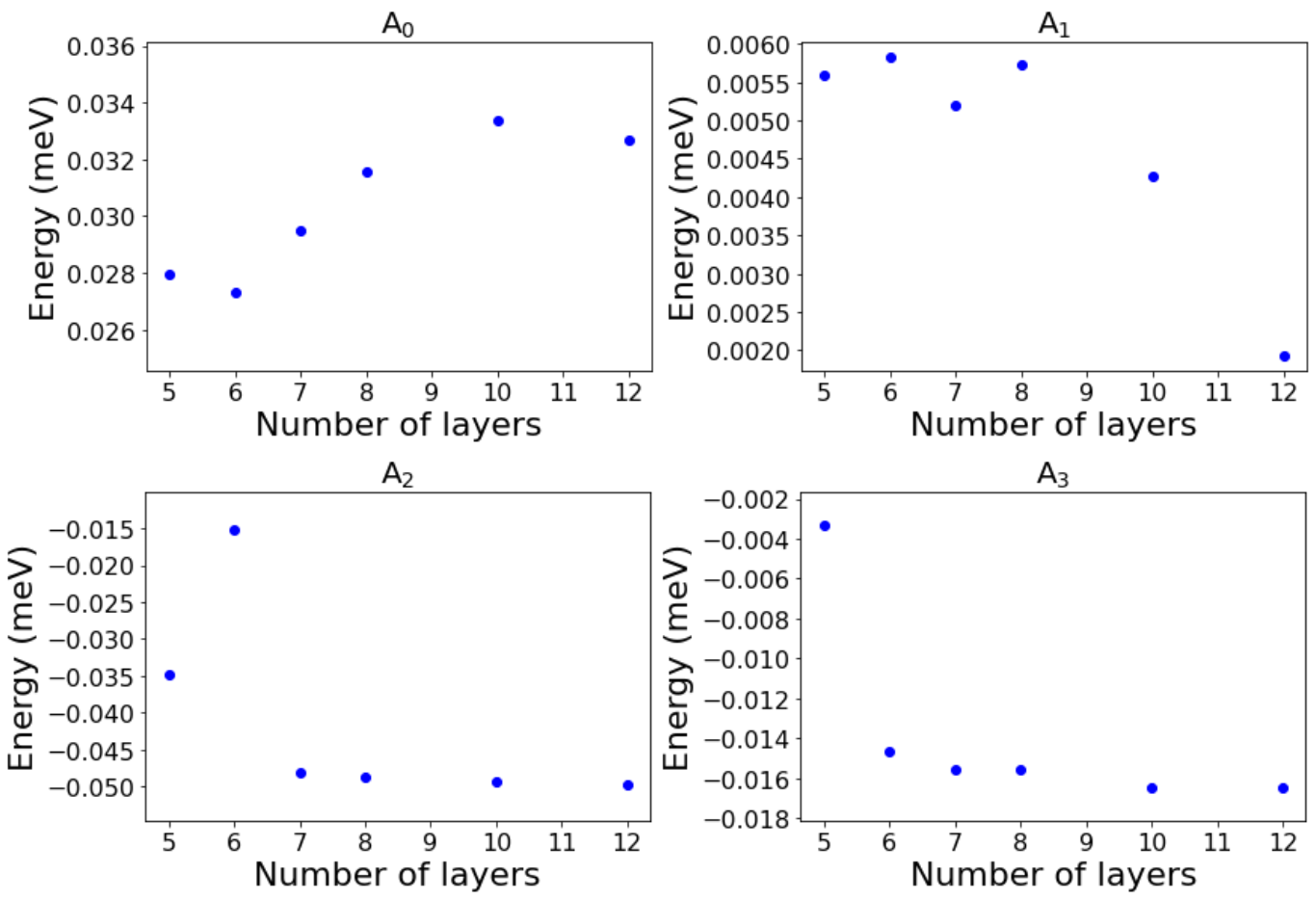}}
\caption{Change of the coefficients with $N$. $A_1 << A_2$ for all $N$, and goes to 0 for larger $N$ (favoring BG). $A_2$ and $A_3$ are relatively constant for $N$ as small as 6. This shows why the $N=12$ fit works well for smaller $N$ (see Fig.~\ref{fig:fit_smaller_N}), and why it works for larger $N$ as well (see Fig.~\ref{fig:N=20}).}
\label{fig:coeffs_vs_N}
\end{figure*}

In Fig.~\ref{fig:N=20}, the coefficients of the $N=12$ Ising fit are used to calculate the energy of mixed $N=20$ BG-RG configurations with a soft interface, and compared to the DFT results. The agreement is very good. Also, Fig.~\ref{fig:fit_smaller_N} shows the $N=12$ fit applied to lower $N$, which also works very well, specially for $N \geq 7$. This is not surprising, considering how the coefficients change with $N$ (Fig.~\ref{fig:coeffs_vs_N}). $A_0$ does change a little from $N=6$ to $N=12$, explaining why the fits are slightly shifted. $A_1$ gets smaller and smaller for larger $N$, and both $A_2$ and $A_3$ barely change beyond $N=7$. Lastly, in the bulk limit, the energy difference between RG and BG is $A_1 + A_2 = -0.015$, although the actual value is around -0.025. The differences in Fig.~\ref{fig:N=20} for $N=20$ are smaller, but the fit cannot be fully extended to the bulk limit (which actually requires larger $N$, due to the significant difference in surface energy).

\begin{table}
{\begin{tabular}{cccc}
\hline \hline
 & (i) $c_d/a = 1.36$ &  (ii) $c_d/a = 1.33$ & (iii) $c_d/a = 1.31$ \\
\hline
$A_0$ & 0.0327 & 0.0730 & 0.1155\\
\hline \\
$A_1$ & 0.0019 & 0.0332 & 0.0684\\ 
\hline \\
$A_2$ &  -0.0497 & -0.0699 & -0.0874 \\
\hline \\
$A_3$  & -0.0164  & -0.0244 & -0.0317 \\
\hline \hline \\
\end{tabular}} \\
\caption{Ising fit coefficients for $N=12$, at room temperature for the experimental lattice parameter $a=2.461$ and different interlayer distances: (i) $c_d/a=1.36$ (also included in Table~\ref{tab:Ising_fit}), (ii) $c_d/a=1.33$, and (iii) $c_d/a=1.31$ (see Fig.~\ref{fig:c_vs_a}). In (i), $|A_2|$ is big relative to $A_1$ and $A_3$, and the low energy states are BG-RG states with soft interfaces. In (ii), $|A_1/A_2|$ is considerably larger, and soft interfaces are also favored, but the very first states are of the type $+-+-++-+-$, so BG with 2 consecutive signs in some part of the sequence. In (iii), the ratio further increases, and soft interfaces are less favored.}
\label{tab:Ising_fit_roomT}
\end{table}



\begin{table*}[b]

{\begin{tabular}{cccc}
\hline \hline
\multicolumn{1}{l}{} & \multicolumn{1}{l}{} & \multicolumn{2}{c}{Energy (meV/atom)} \\
\multicolumn{1}{l}{} & Form & \begin{tabular}[c]{@{}c@{}} $T=284$ K \end{tabular}  & \begin{tabular}[c]{@{}c@{}} $T=1010$ K \end{tabular} \\
\hline \\
Soft BG-RG & $ +-+-++++$ &  0.07 & 0.03  \\
\hline \\
Hard BG-RG & $ +-++---$ &  0.18 & 0.09  \\
\hline \\
Soft BG-BG & $ +-+--+-+$ &  0.06 & 0.07  \\
\hline \\
Hard BG-BG (SF)   & {\small $+-+---+-+$} & 0.07 & 0.09 \\
\hline \\
Soft RG-RG (SF)  & $+++-++++$	& 0.16 & 0.01 \\
\hline \\
Hard RG-RG   & $++++----$	& 0.20  & 0.05 \\
\hline \hline \\
\end{tabular}} \\
\caption{Energy of soft and hard interfaces in BG-RG, BG-BG and RG-RG systems. SF stands for stacking fault. As stated in the main text, the soft BG-RG interface has much lower energy than the hard interface. At 1010 K, they are both reduced by half. The BG-BG interfaces have an energy similar to the soft BG-RG interface, which do not change much at higher temperature.  Instead, the RG-RG interfaces are very energetic at room temperature (in fact, the highest energy states have hard RG-RG interfaces). At 1010K, the soft $E_\textrm{int,RG-RG}$ becomes negligible, while the hard one is similar to the energy of the other interfaces.}
\label{tab:energies_int}
\end{table*}


\begin{figure}[h]
\centering
	{\includegraphics[width=0.45\textwidth]{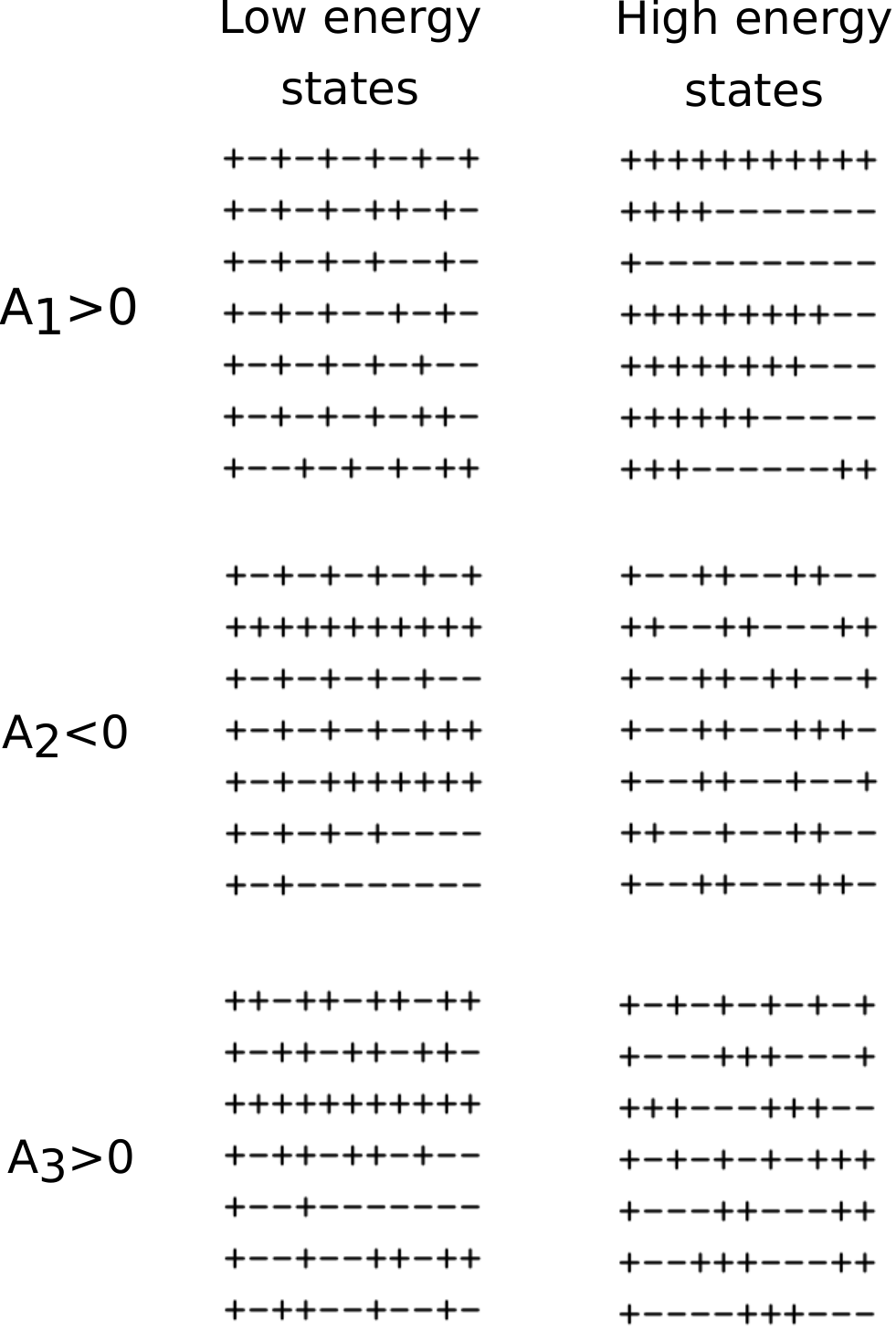}}
\caption{Low and high energy states for each of the coefficients (setting the other ones to 0). $A_1>0$ favors BG, or mostly repeated $+-$ except for some repeated sign. $A_2<0$ favors BG, RG and soft interfaces, while $A_3>0$ favors RG and states of the type $++-++-++-$. The text in the Supporting Information analyzes how this coefficients combine to give the low energy states.}
\label{fig:c_vs_a_GGA}
\end{figure}

\begin{figure}[h]
\centering
	{\includegraphics[width=0.45\textwidth]{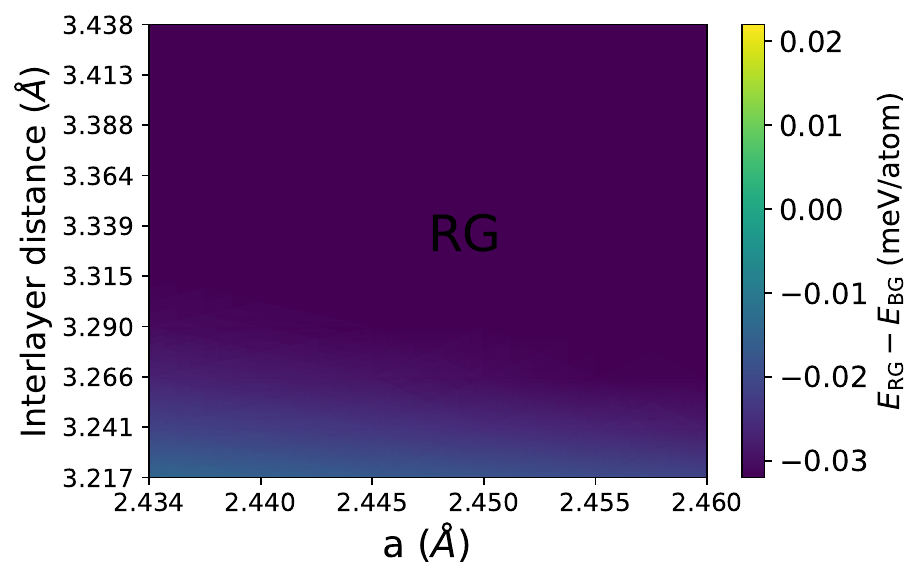}}
\caption{Bulk GGA-PBE energy difference as a function of $c_d$ (interlayer distance) and $a$ at room temperature. Colors are the same as in Fig.~\ref{fig:c_vs_a}. Here, RG is always favored.}
\label{fig:c_vs_a_GGA}
\end{figure}

\begin{figure}[h]
\centering
	{\includegraphics[width=0.45\textwidth]{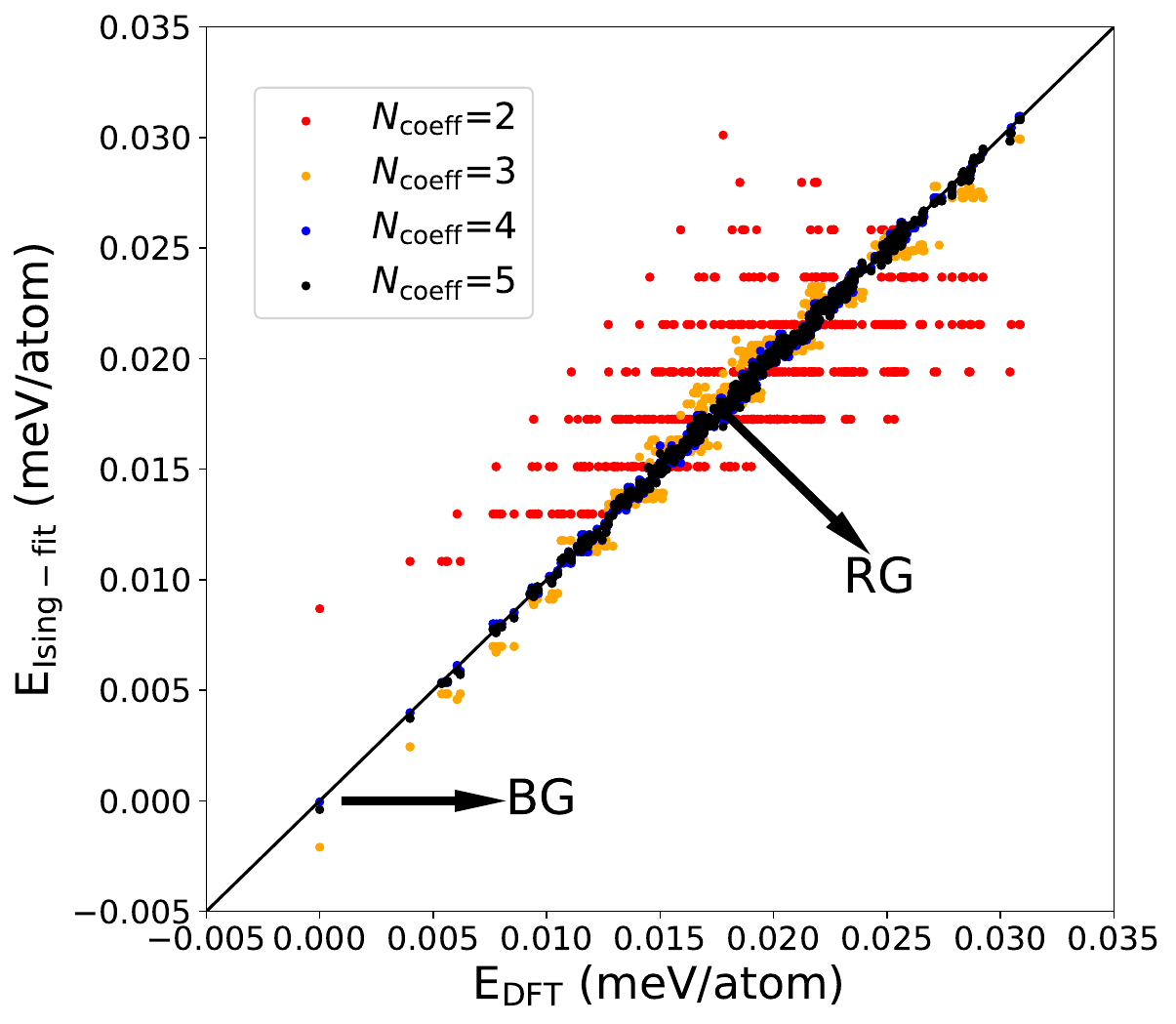}}
\caption{Figure analogous to Fig.~\ref{fig:fit_coeffs}, for $T=1010$ K. In this case, RG falls in the middle of the spectrum (it is not one of the low energy states), and a fit using $A_0$ and $A_1$ works better compared to $T = 284$ K. The fit with $A_i,i=0,1,2,3$ works very well, with all states falling very close to the diagonal. The values of the coefficients are in Table~\ref{tab:Ising_fit}.}
\label{fig:fit_coeffs_T1010}
\end{figure}

\clearpage

\bibliography{bibliography_energies}

\begin{thebibliography}{44}%
\makeatletter
\providecommand \@ifxundefined [1]{%
 \@ifx{#1\undefined}
}%
\providecommand \@ifnum [1]{%
 \ifnum #1\expandafter \@firstoftwo
 \else \expandafter \@secondoftwo
 \fi
}%
\providecommand \@ifx [1]{%
 \ifx #1\expandafter \@firstoftwo
 \else \expandafter \@secondoftwo
 \fi
}%
\providecommand \natexlab [1]{#1}%
\providecommand \enquote  [1]{``#1''}%
\providecommand \bibnamefont  [1]{#1}%
\providecommand \bibfnamefont [1]{#1}%
\providecommand \citenamefont [1]{#1}%
\providecommand \href@noop [0]{\@secondoftwo}%
\providecommand \href [0]{\begingroup \@sanitize@url \@href}%
\providecommand \@href[1]{\@@startlink{#1}\@@href}%
\providecommand \@@href[1]{\endgroup#1\@@endlink}%
\providecommand \@sanitize@url [0]{\catcode `\\12\catcode `\$12\catcode
  `\&12\catcode `\#12\catcode `\^12\catcode `\_12\catcode `\%12\relax}%
\providecommand \@@startlink[1]{}%
\providecommand \@@endlink[0]{}%
\providecommand \url  [0]{\begingroup\@sanitize@url \@url }%
\providecommand \@url [1]{\endgroup\@href {#1}{\urlprefix }}%
\providecommand \urlprefix  [0]{URL }%
\providecommand \Eprint [0]{\href }%
\providecommand \doibase [0]{http://dx.doi.org/}%
\providecommand \selectlanguage [0]{\@gobble}%
\providecommand \bibinfo  [0]{\@secondoftwo}%
\providecommand \bibfield  [0]{\@secondoftwo}%
\providecommand \translation [1]{[#1]}%
\providecommand \BibitemOpen [0]{}%
\providecommand \bibitemStop [0]{}%
\providecommand \bibitemNoStop [0]{.\EOS\space}%
\providecommand \EOS [0]{\spacefactor3000\relax}%
\providecommand \BibitemShut  [1]{\csname bibitem#1\endcsname}%
\let\auto@bib@innerbib\@empty
\bibitem [{Lip(1942)}]{Lipson1942}%
  \BibitemOpen
  \href {\doibase 10.1098/rspa.1942.0063} {\bibfield  {journal} {\bibinfo
  {journal} {Proceedings of the Royal Society of London. Series A. Mathematical
  and Physical Sciences}\ }\textbf {\bibinfo {volume} {181}},\ \bibinfo {pages}
  {101} (\bibinfo {year} {1942})}\BibitemShut {NoStop}%
\bibitem [{\citenamefont {Pierucci}\ \emph {et~al.}(2015)\citenamefont
  {Pierucci}, \citenamefont {Sediri}, \citenamefont {Hajlaoui}, \citenamefont
  {Girard}, \citenamefont {Brumme}, \citenamefont {Calandra}, \citenamefont
  {Velez-Fort}, \citenamefont {Patriarche}, \citenamefont {Silly},
  \citenamefont {Ferro}, \citenamefont {Souli{\`{e}}re}, \citenamefont
  {Marangolo}, \citenamefont {Sirotti}, \citenamefont {Mauri},\ and\
  \citenamefont {Ouerghi}}]{Pierucci2015}%
  \BibitemOpen
  \bibfield  {author} {\bibinfo {author} {\bibfnamefont {D.}~\bibnamefont
  {Pierucci}}, \bibinfo {author} {\bibfnamefont {H.}~\bibnamefont {Sediri}},
  \bibinfo {author} {\bibfnamefont {M.}~\bibnamefont {Hajlaoui}}, \bibinfo
  {author} {\bibfnamefont {J.-C.}\ \bibnamefont {Girard}}, \bibinfo {author}
  {\bibfnamefont {T.}~\bibnamefont {Brumme}}, \bibinfo {author} {\bibfnamefont
  {M.}~\bibnamefont {Calandra}}, \bibinfo {author} {\bibfnamefont
  {E.}~\bibnamefont {Velez-Fort}}, \bibinfo {author} {\bibfnamefont
  {G.}~\bibnamefont {Patriarche}}, \bibinfo {author} {\bibfnamefont {M.~G.}\
  \bibnamefont {Silly}}, \bibinfo {author} {\bibfnamefont {G.}~\bibnamefont
  {Ferro}}, \bibinfo {author} {\bibfnamefont {V.}~\bibnamefont
  {Souli{\`{e}}re}}, \bibinfo {author} {\bibfnamefont {M.}~\bibnamefont
  {Marangolo}}, \bibinfo {author} {\bibfnamefont {F.}~\bibnamefont {Sirotti}},
  \bibinfo {author} {\bibfnamefont {F.}~\bibnamefont {Mauri}}, \ and\ \bibinfo
  {author} {\bibfnamefont {A.}~\bibnamefont {Ouerghi}},\ }\href {\doibase
  10.1021/acsnano.5b01239} {\bibfield  {journal} {\bibinfo  {journal} {{ACS}
  Nano}\ }\textbf {\bibinfo {volume} {9}},\ \bibinfo {pages} {5432} (\bibinfo
  {year} {2015})}\BibitemShut {NoStop}%
\bibitem [{\citenamefont {Henck}\ \emph {et~al.}(2018)\citenamefont {Henck},
  \citenamefont {Avila}, \citenamefont {Aziza}, \citenamefont {Pierucci},
  \citenamefont {Baima}, \citenamefont {Pamuk}, \citenamefont {Chaste},
  \citenamefont {Utt}, \citenamefont {Bartos}, \citenamefont {Nogajewski},
  \citenamefont {Piot}, \citenamefont {Orlita}, \citenamefont {Potemski},
  \citenamefont {Calandra}, \citenamefont {Asensio}, \citenamefont {Mauri},
  \citenamefont {Faugeras},\ and\ \citenamefont {Ouerghi}}]{Henck2018}%
  \BibitemOpen
  \bibfield  {author} {\bibinfo {author} {\bibfnamefont {H.}~\bibnamefont
  {Henck}}, \bibinfo {author} {\bibfnamefont {J.}~\bibnamefont {Avila}},
  \bibinfo {author} {\bibfnamefont {Z.~B.}\ \bibnamefont {Aziza}}, \bibinfo
  {author} {\bibfnamefont {D.}~\bibnamefont {Pierucci}}, \bibinfo {author}
  {\bibfnamefont {J.}~\bibnamefont {Baima}}, \bibinfo {author} {\bibfnamefont
  {B.}~\bibnamefont {Pamuk}}, \bibinfo {author} {\bibfnamefont
  {J.}~\bibnamefont {Chaste}}, \bibinfo {author} {\bibfnamefont
  {D.}~\bibnamefont {Utt}}, \bibinfo {author} {\bibfnamefont {M.}~\bibnamefont
  {Bartos}}, \bibinfo {author} {\bibfnamefont {K.}~\bibnamefont {Nogajewski}},
  \bibinfo {author} {\bibfnamefont {B.~A.}\ \bibnamefont {Piot}}, \bibinfo
  {author} {\bibfnamefont {M.}~\bibnamefont {Orlita}}, \bibinfo {author}
  {\bibfnamefont {M.}~\bibnamefont {Potemski}}, \bibinfo {author}
  {\bibfnamefont {M.}~\bibnamefont {Calandra}}, \bibinfo {author}
  {\bibfnamefont {M.~C.}\ \bibnamefont {Asensio}}, \bibinfo {author}
  {\bibfnamefont {F.}~\bibnamefont {Mauri}}, \bibinfo {author} {\bibfnamefont
  {C.}~\bibnamefont {Faugeras}}, \ and\ \bibinfo {author} {\bibfnamefont
  {A.}~\bibnamefont {Ouerghi}},\ }\href {\doibase 10.1103/physrevb.97.245421}
  {\bibfield  {journal} {\bibinfo  {journal} {Physical Review B}\ }\textbf
  {\bibinfo {volume} {97}} (\bibinfo {year} {2018}),\
  10.1103/physrevb.97.245421}\BibitemShut {NoStop}%
\bibitem [{\citenamefont {Koshino}(2010)}]{Koshino2010}%
  \BibitemOpen
  \bibfield  {author} {\bibinfo {author} {\bibfnamefont {M.}~\bibnamefont
  {Koshino}},\ }\href {\doibase 10.1103/physrevb.81.125304} {\bibfield
  {journal} {\bibinfo  {journal} {Physical Review B}\ }\textbf {\bibinfo
  {volume} {81}} (\bibinfo {year} {2010}),\
  10.1103/physrevb.81.125304}\BibitemShut {NoStop}%
\bibitem [{\citenamefont {Xiao}\ \emph {et~al.}(2011)\citenamefont {Xiao},
  \citenamefont {Tasn{\'{a}}di}, \citenamefont {Koepernik}, \citenamefont
  {Venderbos}, \citenamefont {Richter},\ and\ \citenamefont {Taut}}]{Xiao2011}%
  \BibitemOpen
  \bibfield  {author} {\bibinfo {author} {\bibfnamefont {R.}~\bibnamefont
  {Xiao}}, \bibinfo {author} {\bibfnamefont {F.}~\bibnamefont {Tasn{\'{a}}di}},
  \bibinfo {author} {\bibfnamefont {K.}~\bibnamefont {Koepernik}}, \bibinfo
  {author} {\bibfnamefont {J.~W.~F.}\ \bibnamefont {Venderbos}}, \bibinfo
  {author} {\bibfnamefont {M.}~\bibnamefont {Richter}}, \ and\ \bibinfo
  {author} {\bibfnamefont {M.}~\bibnamefont {Taut}},\ }\href {\doibase
  10.1103/physrevb.84.165404} {\bibfield  {journal} {\bibinfo  {journal}
  {Physical Review B}\ }\textbf {\bibinfo {volume} {84}} (\bibinfo {year}
  {2011}),\ 10.1103/physrevb.84.165404}\BibitemShut {NoStop}%
\bibitem [{\citenamefont {Kopnin}\ \emph {et~al.}(2013)\citenamefont {Kopnin},
  \citenamefont {Ij\"{a}s}, \citenamefont {Harju},\ and\ \citenamefont
  {Heikkil\"{a}}}]{Kopnin2013}%
  \BibitemOpen
  \bibfield  {author} {\bibinfo {author} {\bibfnamefont {N.~B.}\ \bibnamefont
  {Kopnin}}, \bibinfo {author} {\bibfnamefont {M.}~\bibnamefont {Ij\"{a}s}},
  \bibinfo {author} {\bibfnamefont {A.}~\bibnamefont {Harju}}, \ and\ \bibinfo
  {author} {\bibfnamefont {T.~T.}\ \bibnamefont {Heikkil\"{a}}},\ }\href
  {\doibase 10.1103/physrevb.87.140503} {\bibfield  {journal} {\bibinfo
  {journal} {Physical Review B}\ }\textbf {\bibinfo {volume} {87}} (\bibinfo
  {year} {2013}),\ 10.1103/physrevb.87.140503}\BibitemShut {NoStop}%
\bibitem [{\citenamefont {Mu{\~{n}}oz}\ \emph {et~al.}(2013)\citenamefont
  {Mu{\~{n}}oz}, \citenamefont {Covaci},\ and\ \citenamefont
  {Peeters}}]{Munioz2013}%
  \BibitemOpen
  \bibfield  {author} {\bibinfo {author} {\bibfnamefont {W.~A.}\ \bibnamefont
  {Mu{\~{n}}oz}}, \bibinfo {author} {\bibfnamefont {L.}~\bibnamefont {Covaci}},
  \ and\ \bibinfo {author} {\bibfnamefont {F.~M.}\ \bibnamefont {Peeters}},\
  }\href {\doibase 10.1103/physrevb.87.134509} {\bibfield  {journal} {\bibinfo
  {journal} {Physical Review B}\ }\textbf {\bibinfo {volume} {87}} (\bibinfo
  {year} {2013}),\ 10.1103/physrevb.87.134509}\BibitemShut {NoStop}%
\bibitem [{\citenamefont {Otani}\ \emph {et~al.}(2010)\citenamefont {Otani},
  \citenamefont {Koshino}, \citenamefont {Takagi},\ and\ \citenamefont
  {Okada}}]{Otani2010}%
  \BibitemOpen
  \bibfield  {author} {\bibinfo {author} {\bibfnamefont {M.}~\bibnamefont
  {Otani}}, \bibinfo {author} {\bibfnamefont {M.}~\bibnamefont {Koshino}},
  \bibinfo {author} {\bibfnamefont {Y.}~\bibnamefont {Takagi}}, \ and\ \bibinfo
  {author} {\bibfnamefont {S.}~\bibnamefont {Okada}},\ }\href {\doibase
  10.1103/physrevb.81.161403} {\bibfield  {journal} {\bibinfo  {journal}
  {Physical Review B}\ }\textbf {\bibinfo {volume} {81}} (\bibinfo {year}
  {2010}),\ 10.1103/physrevb.81.161403}\BibitemShut {NoStop}%
\bibitem [{\citenamefont {Pamuk}\ \emph {et~al.}(2017)\citenamefont {Pamuk},
  \citenamefont {Baima}, \citenamefont {Mauri},\ and\ \citenamefont
  {Calandra}}]{Pamuk2017}%
  \BibitemOpen
  \bibfield  {author} {\bibinfo {author} {\bibfnamefont {B.}~\bibnamefont
  {Pamuk}}, \bibinfo {author} {\bibfnamefont {J.}~\bibnamefont {Baima}},
  \bibinfo {author} {\bibfnamefont {F.}~\bibnamefont {Mauri}}, \ and\ \bibinfo
  {author} {\bibfnamefont {M.}~\bibnamefont {Calandra}},\ }\href {\doibase
  10.1103/physrevb.95.075422} {\bibfield  {journal} {\bibinfo  {journal}
  {Physical Review B}\ }\textbf {\bibinfo {volume} {95}} (\bibinfo {year}
  {2017}),\ 10.1103/physrevb.95.075422}\BibitemShut {NoStop}%
\bibitem [{\citenamefont {Baima}\ \emph {et~al.}(2018)\citenamefont {Baima},
  \citenamefont {Mauri},\ and\ \citenamefont {Calandra}}]{Baima2018}%
  \BibitemOpen
  \bibfield  {author} {\bibinfo {author} {\bibfnamefont {J.}~\bibnamefont
  {Baima}}, \bibinfo {author} {\bibfnamefont {F.}~\bibnamefont {Mauri}}, \ and\
  \bibinfo {author} {\bibfnamefont {M.}~\bibnamefont {Calandra}},\ }\href
  {\doibase 10.1103/physrevb.98.075418} {\bibfield  {journal} {\bibinfo
  {journal} {Physical Review B}\ }\textbf {\bibinfo {volume} {98}} (\bibinfo
  {year} {2018}),\ 10.1103/physrevb.98.075418}\BibitemShut {NoStop}%
\bibitem [{\citenamefont {Precker}\ \emph {et~al.}(2016)\citenamefont
  {Precker}, \citenamefont {Esquinazi}, \citenamefont {Champi}, \citenamefont
  {Barzola-Quiquia}, \citenamefont {Zoraghi}, \citenamefont
  {Mui{\~{n}}os-Landin}, \citenamefont {Setzer}, \citenamefont {B\"{o}hlmann},
  \citenamefont {Spemann}, \citenamefont {Meijer}, \citenamefont {Muenster},
  \citenamefont {Baehre}, \citenamefont {Kloess},\ and\ \citenamefont
  {Beth}}]{Precker2016}%
  \BibitemOpen
  \bibfield  {author} {\bibinfo {author} {\bibfnamefont {C.~E.}\ \bibnamefont
  {Precker}}, \bibinfo {author} {\bibfnamefont {P.~D.}\ \bibnamefont
  {Esquinazi}}, \bibinfo {author} {\bibfnamefont {A.}~\bibnamefont {Champi}},
  \bibinfo {author} {\bibfnamefont {J.}~\bibnamefont {Barzola-Quiquia}},
  \bibinfo {author} {\bibfnamefont {M.}~\bibnamefont {Zoraghi}}, \bibinfo
  {author} {\bibfnamefont {S.}~\bibnamefont {Mui{\~{n}}os-Landin}}, \bibinfo
  {author} {\bibfnamefont {A.}~\bibnamefont {Setzer}}, \bibinfo {author}
  {\bibfnamefont {W.}~\bibnamefont {B\"{o}hlmann}}, \bibinfo {author}
  {\bibfnamefont {D.}~\bibnamefont {Spemann}}, \bibinfo {author} {\bibfnamefont
  {J.}~\bibnamefont {Meijer}}, \bibinfo {author} {\bibfnamefont
  {T.}~\bibnamefont {Muenster}}, \bibinfo {author} {\bibfnamefont
  {O.}~\bibnamefont {Baehre}}, \bibinfo {author} {\bibfnamefont
  {G.}~\bibnamefont {Kloess}}, \ and\ \bibinfo {author} {\bibfnamefont
  {H.}~\bibnamefont {Beth}},\ }\href {\doibase 10.1088/1367-2630/18/11/113041}
  {\bibfield  {journal} {\bibinfo  {journal} {New Journal of Physics}\ }\textbf
  {\bibinfo {volume} {18}},\ \bibinfo {pages} {113041} (\bibinfo {year}
  {2016})}\BibitemShut {NoStop}%
\bibitem [{\citenamefont {Ballestar}\ \emph {et~al.}(2013)\citenamefont
  {Ballestar}, \citenamefont {Barzola-Quiquia}, \citenamefont {Scheike},\ and\
  \citenamefont {Esquinazi}}]{Ballestar2013}%
  \BibitemOpen
  \bibfield  {author} {\bibinfo {author} {\bibfnamefont {A.}~\bibnamefont
  {Ballestar}}, \bibinfo {author} {\bibfnamefont {J.}~\bibnamefont
  {Barzola-Quiquia}}, \bibinfo {author} {\bibfnamefont {T.}~\bibnamefont
  {Scheike}}, \ and\ \bibinfo {author} {\bibfnamefont {P.}~\bibnamefont
  {Esquinazi}},\ }\href {\doibase 10.1088/1367-2630/15/2/023024} {\bibfield
  {journal} {\bibinfo  {journal} {New Journal of Physics}\ }\textbf {\bibinfo
  {volume} {15}},\ \bibinfo {pages} {023024} (\bibinfo {year}
  {2013})}\BibitemShut {NoStop}%
\bibitem [{\citenamefont {Laves}\ and\ \citenamefont
  {Baskin}(1956)}]{Laves1956}%
  \BibitemOpen
  \bibfield  {author} {\bibinfo {author} {\bibfnamefont {F.}~\bibnamefont
  {Laves}}\ and\ \bibinfo {author} {\bibfnamefont {Y.}~\bibnamefont {Baskin}},\
  }\href {\doibase 10.1524/zkri.1956.107.5-6.337} {\bibfield  {journal}
  {\bibinfo  {journal} {Zeitschrift f\"{u}r Kristallographie}\ }\textbf
  {\bibinfo {volume} {107}},\ \bibinfo {pages} {337} (\bibinfo {year}
  {1956})}\BibitemShut {NoStop}%
\bibitem [{\citenamefont {Boehm}\ and\ \citenamefont
  {Coughlin}(1964)}]{Boehm1964}%
  \BibitemOpen
  \bibfield  {author} {\bibinfo {author} {\bibfnamefont {H.}~\bibnamefont
  {Boehm}}\ and\ \bibinfo {author} {\bibfnamefont {R.}~\bibnamefont
  {Coughlin}},\ }\href {\doibase 10.1016/0008-6223(64)90022-3} {\bibfield
  {journal} {\bibinfo  {journal} {Carbon}\ }\textbf {\bibinfo {volume} {2}},\
  \bibinfo {pages} {1} (\bibinfo {year} {1964})}\BibitemShut {NoStop}%
\bibitem [{\citenamefont {Nery}\ \emph {et~al.}(2020)\citenamefont {Nery},
  \citenamefont {Calandra},\ and\ \citenamefont {Mauri}}]{Nery2020}%
  \BibitemOpen
  \bibfield  {author} {\bibinfo {author} {\bibfnamefont {J.~P.}\ \bibnamefont
  {Nery}}, \bibinfo {author} {\bibfnamefont {M.}~\bibnamefont {Calandra}}, \
  and\ \bibinfo {author} {\bibfnamefont {F.}~\bibnamefont {Mauri}},\ }\href
  {\doibase 10.1021/acs.nanolett.0c01146} {\bibfield  {journal} {\bibinfo
  {journal} {Nano Letters}\ }\textbf {\bibinfo {volume} {20}},\ \bibinfo
  {pages} {5017} (\bibinfo {year} {2020})}\BibitemShut {NoStop}%
\bibitem [{\citenamefont {Lin}\ \emph {et~al.}(2012)\citenamefont {Lin},
  \citenamefont {Li}, \citenamefont {Liu}, \citenamefont {Song}, \citenamefont
  {He}, \citenamefont {Hu}, \citenamefont {Guo},\ and\ \citenamefont
  {Ye}}]{Lin2012}%
  \BibitemOpen
  \bibfield  {author} {\bibinfo {author} {\bibfnamefont {Q.}~\bibnamefont
  {Lin}}, \bibinfo {author} {\bibfnamefont {T.}~\bibnamefont {Li}}, \bibinfo
  {author} {\bibfnamefont {Z.}~\bibnamefont {Liu}}, \bibinfo {author}
  {\bibfnamefont {Y.}~\bibnamefont {Song}}, \bibinfo {author} {\bibfnamefont
  {L.}~\bibnamefont {He}}, \bibinfo {author} {\bibfnamefont {Z.}~\bibnamefont
  {Hu}}, \bibinfo {author} {\bibfnamefont {Q.}~\bibnamefont {Guo}}, \ and\
  \bibinfo {author} {\bibfnamefont {H.}~\bibnamefont {Ye}},\ }\href {\doibase
  10.1016/j.carbon.2012.01.054} {\bibfield  {journal} {\bibinfo  {journal}
  {Carbon}\ }\textbf {\bibinfo {volume} {50}},\ \bibinfo {pages} {2369}
  (\bibinfo {year} {2012})}\BibitemShut {NoStop}%
\bibitem [{\citenamefont {Henni}\ \emph {et~al.}(2016)\citenamefont {Henni},
  \citenamefont {Collado}, \citenamefont {Nogajewski}, \citenamefont {Molas},
  \citenamefont {Usaj}, \citenamefont {Balseiro}, \citenamefont {Orlita},
  \citenamefont {Potemski},\ and\ \citenamefont {Faugeras}}]{Henni2016}%
  \BibitemOpen
  \bibfield  {author} {\bibinfo {author} {\bibfnamefont {Y.}~\bibnamefont
  {Henni}}, \bibinfo {author} {\bibfnamefont {H.~P.~O.}\ \bibnamefont
  {Collado}}, \bibinfo {author} {\bibfnamefont {K.}~\bibnamefont {Nogajewski}},
  \bibinfo {author} {\bibfnamefont {M.~R.}\ \bibnamefont {Molas}}, \bibinfo
  {author} {\bibfnamefont {G.}~\bibnamefont {Usaj}}, \bibinfo {author}
  {\bibfnamefont {C.~A.}\ \bibnamefont {Balseiro}}, \bibinfo {author}
  {\bibfnamefont {M.}~\bibnamefont {Orlita}}, \bibinfo {author} {\bibfnamefont
  {M.}~\bibnamefont {Potemski}}, \ and\ \bibinfo {author} {\bibfnamefont
  {C.}~\bibnamefont {Faugeras}},\ }\href {\doibase
  10.1021/acs.nanolett.6b01041} {\bibfield  {journal} {\bibinfo  {journal}
  {Nano Letters}\ }\textbf {\bibinfo {volume} {16}},\ \bibinfo {pages} {3710}
  (\bibinfo {year} {2016})}\BibitemShut {NoStop}%
\bibitem [{\citenamefont {Yang}\ \emph {et~al.}(2019)\citenamefont {Yang},
  \citenamefont {Zou}, \citenamefont {Woods}, \citenamefont {Shi},
  \citenamefont {Yin}, \citenamefont {Xu}, \citenamefont {Ozdemir},
  \citenamefont {Taniguchi}, \citenamefont {Watanabe}, \citenamefont {Geim},
  \citenamefont {Novoselov}, \citenamefont {Haigh},\ and\ \citenamefont
  {Mishchenko}}]{Yang2019}%
  \BibitemOpen
  \bibfield  {author} {\bibinfo {author} {\bibfnamefont {Y.}~\bibnamefont
  {Yang}}, \bibinfo {author} {\bibfnamefont {Y.-C.}\ \bibnamefont {Zou}},
  \bibinfo {author} {\bibfnamefont {C.~R.}\ \bibnamefont {Woods}}, \bibinfo
  {author} {\bibfnamefont {Y.}~\bibnamefont {Shi}}, \bibinfo {author}
  {\bibfnamefont {J.}~\bibnamefont {Yin}}, \bibinfo {author} {\bibfnamefont
  {S.}~\bibnamefont {Xu}}, \bibinfo {author} {\bibfnamefont {S.}~\bibnamefont
  {Ozdemir}}, \bibinfo {author} {\bibfnamefont {T.}~\bibnamefont {Taniguchi}},
  \bibinfo {author} {\bibfnamefont {K.}~\bibnamefont {Watanabe}}, \bibinfo
  {author} {\bibfnamefont {A.~K.}\ \bibnamefont {Geim}}, \bibinfo {author}
  {\bibfnamefont {K.~S.}\ \bibnamefont {Novoselov}}, \bibinfo {author}
  {\bibfnamefont {S.~J.}\ \bibnamefont {Haigh}}, \ and\ \bibinfo {author}
  {\bibfnamefont {A.}~\bibnamefont {Mishchenko}},\ }\href {\doibase
  10.1021/acs.nanolett.9b03014} {\bibfield  {journal} {\bibinfo  {journal}
  {Nano Letters}\ }\textbf {\bibinfo {volume} {19}},\ \bibinfo {pages} {8526}
  (\bibinfo {year} {2019})}\BibitemShut {NoStop}%
\bibitem [{\citenamefont {Bouhafs}\ \emph {et~al.}(2020)\citenamefont
  {Bouhafs}, \citenamefont {Pezzini}, \citenamefont {Mishra}, \citenamefont
  {Mišeikis}, \citenamefont {Niu}, \citenamefont {Struzzi}, \citenamefont
  {Zakharov}, \citenamefont {Forti},\ and\ \citenamefont
  {Coletti}}]{Chams2020}%
  \BibitemOpen
  \bibfield  {author} {\bibinfo {author} {\bibfnamefont {C.}~\bibnamefont
  {Bouhafs}}, \bibinfo {author} {\bibfnamefont {S.}~\bibnamefont {Pezzini}},
  \bibinfo {author} {\bibfnamefont {N.}~\bibnamefont {Mishra}}, \bibinfo
  {author} {\bibfnamefont {V.}~\bibnamefont {Mišeikis}}, \bibinfo {author}
  {\bibfnamefont {Y.}~\bibnamefont {Niu}}, \bibinfo {author} {\bibfnamefont
  {C.}~\bibnamefont {Struzzi}}, \bibinfo {author} {\bibfnamefont {A.~A.}\
  \bibnamefont {Zakharov}}, \bibinfo {author} {\bibfnamefont {S.}~\bibnamefont
  {Forti}}, \ and\ \bibinfo {author} {\bibfnamefont {C.}~\bibnamefont
  {Coletti}},\ }\href {https://arxiv.org/abs/2006.06667} {\bibfield  {journal}
  {\bibinfo  {journal} {arXiv}\ } (\bibinfo {year} {2020})}\BibitemShut
  {NoStop}%
\bibitem [{\citenamefont {Gao}\ \emph {et~al.}(2020)\citenamefont {Gao},
  \citenamefont {Wang}, \citenamefont {Berry}, \citenamefont {Zhang},
  \citenamefont {Gebhardt}, \citenamefont {Parkin}, \citenamefont {Avila},
  \citenamefont {Yi}, \citenamefont {Chen}, \citenamefont {Hurtado-Parra},
  \citenamefont {Drndi{\'{c}}}, \citenamefont {Rappe}, \citenamefont
  {Srolovitz}, \citenamefont {Kikkawa}, \citenamefont {Luo}, \citenamefont
  {Asensio}, \citenamefont {Wang},\ and\ \citenamefont {Johnson}}]{Gao2020}%
  \BibitemOpen
  \bibfield  {author} {\bibinfo {author} {\bibfnamefont {Z.}~\bibnamefont
  {Gao}}, \bibinfo {author} {\bibfnamefont {S.}~\bibnamefont {Wang}}, \bibinfo
  {author} {\bibfnamefont {J.}~\bibnamefont {Berry}}, \bibinfo {author}
  {\bibfnamefont {Q.}~\bibnamefont {Zhang}}, \bibinfo {author} {\bibfnamefont
  {J.}~\bibnamefont {Gebhardt}}, \bibinfo {author} {\bibfnamefont {W.~M.}\
  \bibnamefont {Parkin}}, \bibinfo {author} {\bibfnamefont {J.}~\bibnamefont
  {Avila}}, \bibinfo {author} {\bibfnamefont {H.}~\bibnamefont {Yi}}, \bibinfo
  {author} {\bibfnamefont {C.}~\bibnamefont {Chen}}, \bibinfo {author}
  {\bibfnamefont {S.}~\bibnamefont {Hurtado-Parra}}, \bibinfo {author}
  {\bibfnamefont {M.}~\bibnamefont {Drndi{\'{c}}}}, \bibinfo {author}
  {\bibfnamefont {A.~M.}\ \bibnamefont {Rappe}}, \bibinfo {author}
  {\bibfnamefont {D.~J.}\ \bibnamefont {Srolovitz}}, \bibinfo {author}
  {\bibfnamefont {J.~M.}\ \bibnamefont {Kikkawa}}, \bibinfo {author}
  {\bibfnamefont {Z.}~\bibnamefont {Luo}}, \bibinfo {author} {\bibfnamefont
  {M.~C.}\ \bibnamefont {Asensio}}, \bibinfo {author} {\bibfnamefont
  {F.}~\bibnamefont {Wang}}, \ and\ \bibinfo {author} {\bibfnamefont
  {A.~T.~C.}\ \bibnamefont {Johnson}},\ }\href {\doibase
  10.1038/s41467-019-14022-3} {\bibfield  {journal} {\bibinfo  {journal}
  {Nature Communications}\ }\textbf {\bibinfo {volume} {11}} (\bibinfo {year}
  {2020}),\ 10.1038/s41467-019-14022-3}\BibitemShut {NoStop}%
\bibitem [{\citenamefont {Kerelsky}\ \emph {et~al.}(2019)\citenamefont
  {Kerelsky}, \citenamefont {Rubio-Verdú}, \citenamefont {Xian}, \citenamefont
  {Kennes}, \citenamefont {M.}, \citenamefont {Halbertal}, \citenamefont
  {Finney}, \citenamefont {Song}, \citenamefont {Turkel}, \citenamefont {Wang},
  \citenamefont {Watanabe}, \citenamefont {Taniguchi}, \citenamefont {Hone},
  \citenamefont {Dean}, \citenamefont {Basov}, \citenamefont {Rubio},\ and\
  \citenamefont {Pasupathy}}]{Kerelsky2019}%
  \BibitemOpen
  \bibfield  {author} {\bibinfo {author} {\bibfnamefont {A.~C.}\ \bibnamefont
  {Kerelsky}}, \bibinfo {author} {\bibnamefont {Rubio-Verdú}}, \bibinfo
  {author} {\bibfnamefont {L.}~\bibnamefont {Xian}}, \bibinfo {author}
  {\bibnamefont {Kennes}}, \bibinfo {author} {\bibfnamefont {D.}~\bibnamefont
  {M.}}, \bibinfo {author} {\bibfnamefont {D.}~\bibnamefont {Halbertal}},
  \bibinfo {author} {\bibfnamefont {N.}~\bibnamefont {Finney}}, \bibinfo
  {author} {\bibfnamefont {L.}~\bibnamefont {Song}}, \bibinfo {author}
  {\bibfnamefont {S.}~\bibnamefont {Turkel}}, \bibinfo {author} {\bibfnamefont
  {L.}~\bibnamefont {Wang}}, \bibinfo {author} {\bibfnamefont {K.}~\bibnamefont
  {Watanabe}}, \bibinfo {author} {\bibfnamefont {T.}~\bibnamefont {Taniguchi}},
  \bibinfo {author} {\bibfnamefont {J.}~\bibnamefont {Hone}}, \bibinfo {author}
  {\bibfnamefont {C.}~\bibnamefont {Dean}}, \bibinfo {author} {\bibfnamefont
  {D.}~\bibnamefont {Basov}}, \bibinfo {author} {\bibfnamefont
  {A.}~\bibnamefont {Rubio}}, \ and\ \bibinfo {author} {\bibfnamefont {A.~N.}\
  \bibnamefont {Pasupathy}},\ }\href {https://arxiv.org/abs/1911.00007}
  {\bibfield  {journal} {\bibinfo  {journal} {arXiv}\ } (\bibinfo {year}
  {2019})}\BibitemShut {NoStop}%
\bibitem [{\citenamefont {Yankowitz}\ \emph {et~al.}(2014)\citenamefont
  {Yankowitz}, \citenamefont {Wang}, \citenamefont {Birdwell}, \citenamefont
  {Chen}, \citenamefont {Watanabe}, \citenamefont {Taniguchi}, \citenamefont
  {Jacquod}, \citenamefont {San-Jose}, \citenamefont {Jarillo-Herrero},\ and\
  \citenamefont {LeRoy}}]{Yankowitz2014}%
  \BibitemOpen
  \bibfield  {author} {\bibinfo {author} {\bibfnamefont {M.}~\bibnamefont
  {Yankowitz}}, \bibinfo {author} {\bibfnamefont {J.~I.-J.}\ \bibnamefont
  {Wang}}, \bibinfo {author} {\bibfnamefont {A.~G.}\ \bibnamefont {Birdwell}},
  \bibinfo {author} {\bibfnamefont {Y.-A.}\ \bibnamefont {Chen}}, \bibinfo
  {author} {\bibfnamefont {K.}~\bibnamefont {Watanabe}}, \bibinfo {author}
  {\bibfnamefont {T.}~\bibnamefont {Taniguchi}}, \bibinfo {author}
  {\bibfnamefont {P.}~\bibnamefont {Jacquod}}, \bibinfo {author} {\bibfnamefont
  {P.}~\bibnamefont {San-Jose}}, \bibinfo {author} {\bibfnamefont
  {P.}~\bibnamefont {Jarillo-Herrero}}, \ and\ \bibinfo {author} {\bibfnamefont
  {B.~J.}\ \bibnamefont {LeRoy}},\ }\href {\doibase 10.1038/nmat3965}
  {\bibfield  {journal} {\bibinfo  {journal} {Nature Materials}\ }\textbf
  {\bibinfo {volume} {13}},\ \bibinfo {pages} {786} (\bibinfo {year}
  {2014})}\BibitemShut {NoStop}%
\bibitem [{\citenamefont {Li}\ \emph {et~al.}(2020)\citenamefont {Li},
  \citenamefont {Utama}, \citenamefont {Wang}, \citenamefont {Zhao},
  \citenamefont {Zhao}, \citenamefont {Xiao}, \citenamefont {Jiang},
  \citenamefont {Jiang}, \citenamefont {Taniguchi}, \citenamefont {Watanabe},
  \citenamefont {Weber-Bargioni}, \citenamefont {Zettl},\ and\ \citenamefont
  {Wang}}]{Li2020}%
  \BibitemOpen
  \bibfield  {author} {\bibinfo {author} {\bibfnamefont {H.}~\bibnamefont
  {Li}}, \bibinfo {author} {\bibfnamefont {M.~I.~B.}\ \bibnamefont {Utama}},
  \bibinfo {author} {\bibfnamefont {S.}~\bibnamefont {Wang}}, \bibinfo {author}
  {\bibfnamefont {W.}~\bibnamefont {Zhao}}, \bibinfo {author} {\bibfnamefont
  {S.}~\bibnamefont {Zhao}}, \bibinfo {author} {\bibfnamefont {X.}~\bibnamefont
  {Xiao}}, \bibinfo {author} {\bibfnamefont {Y.}~\bibnamefont {Jiang}},
  \bibinfo {author} {\bibfnamefont {L.}~\bibnamefont {Jiang}}, \bibinfo
  {author} {\bibfnamefont {T.}~\bibnamefont {Taniguchi}}, \bibinfo {author}
  {\bibfnamefont {K.}~\bibnamefont {Watanabe}}, \bibinfo {author}
  {\bibfnamefont {A.}~\bibnamefont {Weber-Bargioni}}, \bibinfo {author}
  {\bibfnamefont {A.}~\bibnamefont {Zettl}}, \ and\ \bibinfo {author}
  {\bibfnamefont {F.}~\bibnamefont {Wang}},\ }\href {\doibase
  10.1021/acs.nanolett.9b05092} {\bibfield  {journal} {\bibinfo  {journal}
  {Nano Letters}\ }\textbf {\bibinfo {volume} {20}},\ \bibinfo {pages} {3106}
  (\bibinfo {year} {2020})}\BibitemShut {NoStop}%
\bibitem [{\citenamefont {Geisenhof}\ \emph {et~al.}(2019)\citenamefont
  {Geisenhof}, \citenamefont {Winterer}, \citenamefont {Wakolbinger},
  \citenamefont {Gokus}, \citenamefont {Durmaz}, \citenamefont {Priesack},
  \citenamefont {Lenz}, \citenamefont {Keilmann}, \citenamefont {Watanabe},
  \citenamefont {Taniguchi}, \citenamefont {Guerrero-Avil{\'{e}}s},
  \citenamefont {Pelc}, \citenamefont {Ayuela},\ and\ \citenamefont
  {Weitz}}]{Geisenhof2019}%
  \BibitemOpen
  \bibfield  {author} {\bibinfo {author} {\bibfnamefont {F.~R.}\ \bibnamefont
  {Geisenhof}}, \bibinfo {author} {\bibfnamefont {F.}~\bibnamefont {Winterer}},
  \bibinfo {author} {\bibfnamefont {S.}~\bibnamefont {Wakolbinger}}, \bibinfo
  {author} {\bibfnamefont {T.~D.}\ \bibnamefont {Gokus}}, \bibinfo {author}
  {\bibfnamefont {Y.~C.}\ \bibnamefont {Durmaz}}, \bibinfo {author}
  {\bibfnamefont {D.}~\bibnamefont {Priesack}}, \bibinfo {author}
  {\bibfnamefont {J.}~\bibnamefont {Lenz}}, \bibinfo {author} {\bibfnamefont
  {F.}~\bibnamefont {Keilmann}}, \bibinfo {author} {\bibfnamefont
  {K.}~\bibnamefont {Watanabe}}, \bibinfo {author} {\bibfnamefont
  {T.}~\bibnamefont {Taniguchi}}, \bibinfo {author} {\bibfnamefont
  {R.}~\bibnamefont {Guerrero-Avil{\'{e}}s}}, \bibinfo {author} {\bibfnamefont
  {M.}~\bibnamefont {Pelc}}, \bibinfo {author} {\bibfnamefont {A.}~\bibnamefont
  {Ayuela}}, \ and\ \bibinfo {author} {\bibfnamefont {R.~T.}\ \bibnamefont
  {Weitz}},\ }\href {\doibase 10.1021/acsanm.9b01603} {\bibfield  {journal}
  {\bibinfo  {journal} {{ACS} Applied Nano Materials}\ }\textbf {\bibinfo
  {volume} {2}},\ \bibinfo {pages} {6067} (\bibinfo {year} {2019})}\BibitemShut
  {NoStop}%
\bibitem [{\citenamefont {Chen}\ \emph {et~al.}(2019)\citenamefont {Chen},
  \citenamefont {Jiang}, \citenamefont {Wu}, \citenamefont {Lyu}, \citenamefont
  {Li}, \citenamefont {Chittari}, \citenamefont {Watanabe}, \citenamefont
  {Taniguchi}, \citenamefont {Shi}, \citenamefont {Jung}, \citenamefont
  {Zhang},\ and\ \citenamefont {Wang}}]{Chen2019}%
  \BibitemOpen
  \bibfield  {author} {\bibinfo {author} {\bibfnamefont {G.}~\bibnamefont
  {Chen}}, \bibinfo {author} {\bibfnamefont {L.}~\bibnamefont {Jiang}},
  \bibinfo {author} {\bibfnamefont {S.}~\bibnamefont {Wu}}, \bibinfo {author}
  {\bibfnamefont {B.}~\bibnamefont {Lyu}}, \bibinfo {author} {\bibfnamefont
  {H.}~\bibnamefont {Li}}, \bibinfo {author} {\bibfnamefont {B.~L.}\
  \bibnamefont {Chittari}}, \bibinfo {author} {\bibfnamefont {K.}~\bibnamefont
  {Watanabe}}, \bibinfo {author} {\bibfnamefont {T.}~\bibnamefont {Taniguchi}},
  \bibinfo {author} {\bibfnamefont {Z.}~\bibnamefont {Shi}}, \bibinfo {author}
  {\bibfnamefont {J.}~\bibnamefont {Jung}}, \bibinfo {author} {\bibfnamefont
  {Y.}~\bibnamefont {Zhang}}, \ and\ \bibinfo {author} {\bibfnamefont
  {F.}~\bibnamefont {Wang}},\ }\href {\doibase 10.1038/s41567-018-0387-2}
  {\bibfield  {journal} {\bibinfo  {journal} {Nature Physics}\ }\textbf
  {\bibinfo {volume} {15}},\ \bibinfo {pages} {237} (\bibinfo {year}
  {2019})}\BibitemShut {NoStop}%
\bibitem [{\citenamefont {Charlier}\ \emph {et~al.}(1994)\citenamefont
  {Charlier}, \citenamefont {Gonze},\ and\ \citenamefont
  {Michenaud}}]{Charlier1994}%
  \BibitemOpen
  \bibfield  {author} {\bibinfo {author} {\bibfnamefont {J.-C.}\ \bibnamefont
  {Charlier}}, \bibinfo {author} {\bibfnamefont {X.}~\bibnamefont {Gonze}}, \
  and\ \bibinfo {author} {\bibfnamefont {J.-P.}\ \bibnamefont {Michenaud}},\
  }\href {\doibase 10.1016/0008-6223(94)90192-9} {\bibfield  {journal}
  {\bibinfo  {journal} {Carbon}\ }\textbf {\bibinfo {volume} {32}},\ \bibinfo
  {pages} {289} (\bibinfo {year} {1994})}\BibitemShut {NoStop}%
\bibitem [{\citenamefont {Savini}\ \emph {et~al.}(2011)\citenamefont {Savini},
  \citenamefont {Dappe}, \citenamefont {\"{O}berg}, \citenamefont {Charlier},
  \citenamefont {Katsnelson},\ and\ \citenamefont {Fasolino}}]{Savini2011}%
  \BibitemOpen
  \bibfield  {author} {\bibinfo {author} {\bibfnamefont {G.}~\bibnamefont
  {Savini}}, \bibinfo {author} {\bibfnamefont {Y.}~\bibnamefont {Dappe}},
  \bibinfo {author} {\bibfnamefont {S.}~\bibnamefont {\"{O}berg}}, \bibinfo
  {author} {\bibfnamefont {J.-C.}\ \bibnamefont {Charlier}}, \bibinfo {author}
  {\bibfnamefont {M.}~\bibnamefont {Katsnelson}}, \ and\ \bibinfo {author}
  {\bibfnamefont {A.}~\bibnamefont {Fasolino}},\ }\href {\doibase
  10.1016/j.carbon.2010.08.042} {\bibfield  {journal} {\bibinfo  {journal}
  {Carbon}\ }\textbf {\bibinfo {volume} {49}},\ \bibinfo {pages} {62} (\bibinfo
  {year} {2011})}\BibitemShut {NoStop}%
\bibitem [{\citenamefont {Anees}\ \emph {et~al.}(2014)\citenamefont {Anees},
  \citenamefont {Valsakumar}, \citenamefont {Chandra},\ and\ \citenamefont
  {Panigrahi}}]{Anees2014}%
  \BibitemOpen
  \bibfield  {author} {\bibinfo {author} {\bibfnamefont {P.}~\bibnamefont
  {Anees}}, \bibinfo {author} {\bibfnamefont {M.~C.}\ \bibnamefont
  {Valsakumar}}, \bibinfo {author} {\bibfnamefont {S.}~\bibnamefont {Chandra}},
  \ and\ \bibinfo {author} {\bibfnamefont {B.~K.}\ \bibnamefont {Panigrahi}},\
  }\href {\doibase 10.1088/0965-0393/22/3/035016} {\bibfield  {journal}
  {\bibinfo  {journal} {Modelling and Simulation in Materials Science and
  Engineering}\ }\textbf {\bibinfo {volume} {22}},\ \bibinfo {pages} {035016}
  (\bibinfo {year} {2014})}\BibitemShut {NoStop}%
\bibitem [{\citenamefont {Taut}\ \emph {et~al.}(2013)\citenamefont {Taut},
  \citenamefont {Koepernik},\ and\ \citenamefont {Richter}}]{Taut2013}%
  \BibitemOpen
  \bibfield  {author} {\bibinfo {author} {\bibfnamefont {M.}~\bibnamefont
  {Taut}}, \bibinfo {author} {\bibfnamefont {K.}~\bibnamefont {Koepernik}}, \
  and\ \bibinfo {author} {\bibfnamefont {M.}~\bibnamefont {Richter}},\ }\href
  {\doibase 10.1103/physrevb.88.205411} {\bibfield  {journal} {\bibinfo
  {journal} {Physical Review B}\ }\textbf {\bibinfo {volume} {88}} (\bibinfo
  {year} {2013}),\ 10.1103/physrevb.88.205411}\BibitemShut {NoStop}%
\bibitem [{\citenamefont {Halilov}\ \emph {et~al.}(1998)\citenamefont
  {Halilov}, \citenamefont {Perlov}, \citenamefont {Oppeneer}, \citenamefont
  {Yaresko},\ and\ \citenamefont {Antonov}}]{Halilov1998}%
  \BibitemOpen
  \bibfield  {author} {\bibinfo {author} {\bibfnamefont {S.~V.}\ \bibnamefont
  {Halilov}}, \bibinfo {author} {\bibfnamefont {A.~Y.}\ \bibnamefont {Perlov}},
  \bibinfo {author} {\bibfnamefont {P.~M.}\ \bibnamefont {Oppeneer}}, \bibinfo
  {author} {\bibfnamefont {A.~N.}\ \bibnamefont {Yaresko}}, \ and\ \bibinfo
  {author} {\bibfnamefont {V.~N.}\ \bibnamefont {Antonov}},\ }\href {\doibase
  10.1103/physrevb.57.9557} {\bibfield  {journal} {\bibinfo  {journal}
  {Physical Review B}\ }\textbf {\bibinfo {volume} {57}},\ \bibinfo {pages}
  {9557} (\bibinfo {year} {1998})}\BibitemShut {NoStop}%
\bibitem [{\citenamefont {Freise}\ and\ \citenamefont
  {Kelly}(1963)}]{Freise1963}%
  \BibitemOpen
  \bibfield  {author} {\bibinfo {author} {\bibfnamefont {E.~J.}\ \bibnamefont
  {Freise}}\ and\ \bibinfo {author} {\bibfnamefont {A.}~\bibnamefont {Kelly}},\
  }\href {\doibase 10.1080/14786436308207315} {\bibfield  {journal} {\bibinfo
  {journal} {Philosophical Magazine}\ }\textbf {\bibinfo {volume} {8}},\
  \bibinfo {pages} {1519} (\bibinfo {year} {1963})}\BibitemShut {NoStop}%
\bibitem [{\citenamefont {Mounet}\ and\ \citenamefont
  {Marzari}(2005)}]{Mounet2005}%
  \BibitemOpen
  \bibfield  {author} {\bibinfo {author} {\bibfnamefont {N.}~\bibnamefont
  {Mounet}}\ and\ \bibinfo {author} {\bibfnamefont {N.}~\bibnamefont
  {Marzari}},\ }\href {\doibase 10.1103/physrevb.71.205214} {\bibfield
  {journal} {\bibinfo  {journal} {Physical Review B}\ }\textbf {\bibinfo
  {volume} {71}} (\bibinfo {year} {2005}),\
  10.1103/physrevb.71.205214}\BibitemShut {NoStop}%
\bibitem [{\citenamefont {Grimme}(2006)}]{Grimme2006}%
  \BibitemOpen
  \bibfield  {author} {\bibinfo {author} {\bibfnamefont {S.}~\bibnamefont
  {Grimme}},\ }\href {\doibase 10.1002/jcc.20495} {\bibfield  {journal}
  {\bibinfo  {journal} {Journal of Computational Chemistry}\ }\textbf {\bibinfo
  {volume} {27}},\ \bibinfo {pages} {1787} (\bibinfo {year}
  {2006})}\BibitemShut {NoStop}%
\bibitem [{\citenamefont {Perdew}\ \emph {et~al.}(1996)\citenamefont {Perdew},
  \citenamefont {Burke},\ and\ \citenamefont {Ernzerhof}}]{Perdew1996}%
  \BibitemOpen
  \bibfield  {author} {\bibinfo {author} {\bibfnamefont {J.~P.}\ \bibnamefont
  {Perdew}}, \bibinfo {author} {\bibfnamefont {K.}~\bibnamefont {Burke}}, \
  and\ \bibinfo {author} {\bibfnamefont {M.}~\bibnamefont {Ernzerhof}},\ }\href
  {\doibase 10.1103/physrevlett.77.3865} {\bibfield  {journal} {\bibinfo
  {journal} {Physical Review Letters}\ }\textbf {\bibinfo {volume} {77}},\
  \bibinfo {pages} {3865} (\bibinfo {year} {1996})}\BibitemShut {NoStop}%
\bibitem [{\citenamefont {Wang}\ \emph {et~al.}(2015)\citenamefont {Wang},
  \citenamefont {Dai}, \citenamefont {Li}, \citenamefont {Yang}, \citenamefont
  {Srolovitz},\ and\ \citenamefont {Zheng}}]{Wang2015}%
  \BibitemOpen
  \bibfield  {author} {\bibinfo {author} {\bibfnamefont {W.}~\bibnamefont
  {Wang}}, \bibinfo {author} {\bibfnamefont {S.}~\bibnamefont {Dai}}, \bibinfo
  {author} {\bibfnamefont {X.}~\bibnamefont {Li}}, \bibinfo {author}
  {\bibfnamefont {J.}~\bibnamefont {Yang}}, \bibinfo {author} {\bibfnamefont
  {D.~J.}\ \bibnamefont {Srolovitz}}, \ and\ \bibinfo {author} {\bibfnamefont
  {Q.}~\bibnamefont {Zheng}},\ }\href {\doibase 10.1038/ncomms8853} {\bibfield
  {journal} {\bibinfo  {journal} {Nature Communications}\ }\textbf {\bibinfo
  {volume} {6}} (\bibinfo {year} {2015}),\ 10.1038/ncomms8853}\BibitemShut
  {NoStop}%
\bibitem [{\citenamefont {Telling}\ and\ \citenamefont
  {Heggie}(2003)}]{Telling2003}%
  \BibitemOpen
  \bibfield  {author} {\bibinfo {author} {\bibfnamefont {R.~H.}\ \bibnamefont
  {Telling}}\ and\ \bibinfo {author} {\bibfnamefont {M.~I.}\ \bibnamefont
  {Heggie}},\ }\href {\doibase 10.1080/0950083031000137839} {\bibfield
  {journal} {\bibinfo  {journal} {Philosophical Magazine Letters}\ }\textbf
  {\bibinfo {volume} {83}},\ \bibinfo {pages} {411} (\bibinfo {year}
  {2003})}\BibitemShut {NoStop}%
\bibitem [{\citenamefont {Zhou}\ \emph {et~al.}(2015)\citenamefont {Zhou},
  \citenamefont {Han}, \citenamefont {Dai}, \citenamefont {Sun},\ and\
  \citenamefont {Srolovitz}}]{Zhou2015}%
  \BibitemOpen
  \bibfield  {author} {\bibinfo {author} {\bibfnamefont {S.}~\bibnamefont
  {Zhou}}, \bibinfo {author} {\bibfnamefont {J.}~\bibnamefont {Han}}, \bibinfo
  {author} {\bibfnamefont {S.}~\bibnamefont {Dai}}, \bibinfo {author}
  {\bibfnamefont {J.}~\bibnamefont {Sun}}, \ and\ \bibinfo {author}
  {\bibfnamefont {D.~J.}\ \bibnamefont {Srolovitz}},\ }\href {\doibase
  10.1103/physrevb.92.155438} {\bibfield  {journal} {\bibinfo  {journal}
  {Physical Review B}\ }\textbf {\bibinfo {volume} {92}} (\bibinfo {year}
  {2015}),\ 10.1103/physrevb.92.155438}\BibitemShut {NoStop}%
\bibitem [{\citenamefont {Mostaani}\ \emph {et~al.}(2015)\citenamefont
  {Mostaani}, \citenamefont {Drummond},\ and\ \citenamefont
  {Fal'ko}}]{Mostaani2015}%
  \BibitemOpen
  \bibfield  {author} {\bibinfo {author} {\bibfnamefont {E.}~\bibnamefont
  {Mostaani}}, \bibinfo {author} {\bibfnamefont {N.}~\bibnamefont {Drummond}},
  \ and\ \bibinfo {author} {\bibfnamefont {V.}~\bibnamefont {Fal'ko}},\ }\href
  {\doibase 10.1103/physrevlett.115.115501} {\bibfield  {journal} {\bibinfo
  {journal} {Physical Review Letters}\ }\textbf {\bibinfo {volume} {115}}
  (\bibinfo {year} {2015}),\ 10.1103/physrevlett.115.115501}\BibitemShut
  {NoStop}%
\bibitem [{\citenamefont {Nover}\ \emph {et~al.}(2005)\citenamefont {Nover},
  \citenamefont {Stoll},\ and\ \citenamefont {von~der G\"{o}nna}}]{Nover2005}%
  \BibitemOpen
  \bibfield  {author} {\bibinfo {author} {\bibfnamefont {G.}~\bibnamefont
  {Nover}}, \bibinfo {author} {\bibfnamefont {J.~B.}\ \bibnamefont {Stoll}}, \
  and\ \bibinfo {author} {\bibfnamefont {J.}~\bibnamefont {von~der
  G\"{o}nna}},\ }\href {\doibase 10.1111/j.1365-246x.2005.02395.x} {\bibfield
  {journal} {\bibinfo  {journal} {Geophysical Journal International}\ }\textbf
  {\bibinfo {volume} {160}},\ \bibinfo {pages} {1059} (\bibinfo {year}
  {2005})}\BibitemShut {NoStop}%
\bibitem [{\citenamefont {Lui}\ \emph {et~al.}(2011)\citenamefont {Lui},
  \citenamefont {Li}, \citenamefont {Chen}, \citenamefont {Klimov},
  \citenamefont {Brus},\ and\ \citenamefont {Heinz}}]{Lui2011}%
  \BibitemOpen
  \bibfield  {author} {\bibinfo {author} {\bibfnamefont {C.~H.}\ \bibnamefont
  {Lui}}, \bibinfo {author} {\bibfnamefont {Z.}~\bibnamefont {Li}}, \bibinfo
  {author} {\bibfnamefont {Z.}~\bibnamefont {Chen}}, \bibinfo {author}
  {\bibfnamefont {P.~V.}\ \bibnamefont {Klimov}}, \bibinfo {author}
  {\bibfnamefont {L.~E.}\ \bibnamefont {Brus}}, \ and\ \bibinfo {author}
  {\bibfnamefont {T.~F.}\ \bibnamefont {Heinz}},\ }\href {\doibase
  10.1021/nl1032827} {\bibfield  {journal} {\bibinfo  {journal} {Nano Letters}\
  }\textbf {\bibinfo {volume} {11}},\ \bibinfo {pages} {164} (\bibinfo {year}
  {2011})}\BibitemShut {NoStop}%
\bibitem [{\citenamefont {Baker}\ \emph {et~al.}(1961)\citenamefont {Baker},
  \citenamefont {Chou},\ and\ \citenamefont {Kelly}}]{Baker1961}%
  \BibitemOpen
  \bibfield  {author} {\bibinfo {author} {\bibfnamefont {C.}~\bibnamefont
  {Baker}}, \bibinfo {author} {\bibfnamefont {Y.~T.}\ \bibnamefont {Chou}}, \
  and\ \bibinfo {author} {\bibfnamefont {A.}~\bibnamefont {Kelly}},\ }\href
  {\doibase 10.1080/14786436108243381} {\bibfield  {journal} {\bibinfo
  {journal} {Philosophical Magazine}\ }\textbf {\bibinfo {volume} {6}},\
  \bibinfo {pages} {1305} (\bibinfo {year} {1961})}\BibitemShut {NoStop}%
\bibitem [{\citenamefont {S~Amelinckx}(1965)}]{Amelinckx1965}%
  \BibitemOpen
  \bibfield  {author} {\bibinfo {author} {\bibfnamefont {M.~H.}\ \bibnamefont
  {S~Amelinckx}, \bibfnamefont {P~Delavignette}},\ }\href@noop {} {\bibfield
  {journal} {\bibinfo  {journal} {Chem. Phys. Carbon}\ }\textbf {\bibinfo
  {volume} {1}} (\bibinfo {year} {1965})}\BibitemShut {NoStop}%
\bibitem [{\citenamefont {Giannozzi}\ \emph {et~al.}(2017)\citenamefont
  {Giannozzi}, \citenamefont {Andreussi}, \citenamefont {Brumme}, \citenamefont
  {Bunau}, \citenamefont {Nardelli}, \citenamefont {Calandra}, \citenamefont
  {Car}, \citenamefont {Cavazzoni}, \citenamefont {Ceresoli}, \citenamefont
  {Cococcioni}, \citenamefont {Colonna}, \citenamefont {Carnimeo},
  \citenamefont {Corso}, \citenamefont {de~Gironcoli}, \citenamefont {Delugas},
  \citenamefont {DiStasio}, \citenamefont {Ferretti}, \citenamefont {Floris},
  \citenamefont {Fratesi}, \citenamefont {Fugallo}, \citenamefont {Gebauer},
  \citenamefont {Gerstmann}, \citenamefont {Giustino}, \citenamefont {Gorni},
  \citenamefont {Jia}, \citenamefont {Kawamura}, \citenamefont {Ko},
  \citenamefont {Kokalj}, \citenamefont {K\"{u}{\c{c}}\"{u}kbenli},
  \citenamefont {Lazzeri}, \citenamefont {Marsili}, \citenamefont {Marzari},
  \citenamefont {Mauri}, \citenamefont {Nguyen}, \citenamefont {Nguyen},
  \citenamefont {de-la Roza}, \citenamefont {Paulatto}, \citenamefont
  {Ponc{\'{e}}}, \citenamefont {Rocca}, \citenamefont {Sabatini}, \citenamefont
  {Santra}, \citenamefont {Schlipf}, \citenamefont {Seitsonen}, \citenamefont
  {Smogunov}, \citenamefont {Timrov}, \citenamefont {Thonhauser}, \citenamefont
  {Umari}, \citenamefont {Vast}, \citenamefont {Wu},\ and\ \citenamefont
  {Baroni}}]{QE2017}%
  \BibitemOpen
  \bibfield  {author} {\bibinfo {author} {\bibfnamefont {P.}~\bibnamefont
  {Giannozzi}}, \bibinfo {author} {\bibfnamefont {O.}~\bibnamefont
  {Andreussi}}, \bibinfo {author} {\bibfnamefont {T.}~\bibnamefont {Brumme}},
  \bibinfo {author} {\bibfnamefont {O.}~\bibnamefont {Bunau}}, \bibinfo
  {author} {\bibfnamefont {M.~B.}\ \bibnamefont {Nardelli}}, \bibinfo {author}
  {\bibfnamefont {M.}~\bibnamefont {Calandra}}, \bibinfo {author}
  {\bibfnamefont {R.}~\bibnamefont {Car}}, \bibinfo {author} {\bibfnamefont
  {C.}~\bibnamefont {Cavazzoni}}, \bibinfo {author} {\bibfnamefont
  {D.}~\bibnamefont {Ceresoli}}, \bibinfo {author} {\bibfnamefont
  {M.}~\bibnamefont {Cococcioni}}, \bibinfo {author} {\bibfnamefont
  {N.}~\bibnamefont {Colonna}}, \bibinfo {author} {\bibfnamefont
  {I.}~\bibnamefont {Carnimeo}}, \bibinfo {author} {\bibfnamefont {A.~D.}\
  \bibnamefont {Corso}}, \bibinfo {author} {\bibfnamefont {S.}~\bibnamefont
  {de~Gironcoli}}, \bibinfo {author} {\bibfnamefont {P.}~\bibnamefont
  {Delugas}}, \bibinfo {author} {\bibfnamefont {R.~A.}\ \bibnamefont
  {DiStasio}}, \bibinfo {author} {\bibfnamefont {A.}~\bibnamefont {Ferretti}},
  \bibinfo {author} {\bibfnamefont {A.}~\bibnamefont {Floris}}, \bibinfo
  {author} {\bibfnamefont {G.}~\bibnamefont {Fratesi}}, \bibinfo {author}
  {\bibfnamefont {G.}~\bibnamefont {Fugallo}}, \bibinfo {author} {\bibfnamefont
  {R.}~\bibnamefont {Gebauer}}, \bibinfo {author} {\bibfnamefont
  {U.}~\bibnamefont {Gerstmann}}, \bibinfo {author} {\bibfnamefont
  {F.}~\bibnamefont {Giustino}}, \bibinfo {author} {\bibfnamefont
  {T.}~\bibnamefont {Gorni}}, \bibinfo {author} {\bibfnamefont
  {J.}~\bibnamefont {Jia}}, \bibinfo {author} {\bibfnamefont {M.}~\bibnamefont
  {Kawamura}}, \bibinfo {author} {\bibfnamefont {H.-Y.}\ \bibnamefont {Ko}},
  \bibinfo {author} {\bibfnamefont {A.}~\bibnamefont {Kokalj}}, \bibinfo
  {author} {\bibfnamefont {E.}~\bibnamefont {K\"{u}{\c{c}}\"{u}kbenli}},
  \bibinfo {author} {\bibfnamefont {M.}~\bibnamefont {Lazzeri}}, \bibinfo
  {author} {\bibfnamefont {M.}~\bibnamefont {Marsili}}, \bibinfo {author}
  {\bibfnamefont {N.}~\bibnamefont {Marzari}}, \bibinfo {author} {\bibfnamefont
  {F.}~\bibnamefont {Mauri}}, \bibinfo {author} {\bibfnamefont {N.~L.}\
  \bibnamefont {Nguyen}}, \bibinfo {author} {\bibfnamefont {H.-V.}\
  \bibnamefont {Nguyen}}, \bibinfo {author} {\bibfnamefont {A.~O.}\
  \bibnamefont {de-la Roza}}, \bibinfo {author} {\bibfnamefont
  {L.}~\bibnamefont {Paulatto}}, \bibinfo {author} {\bibfnamefont
  {S.}~\bibnamefont {Ponc{\'{e}}}}, \bibinfo {author} {\bibfnamefont
  {D.}~\bibnamefont {Rocca}}, \bibinfo {author} {\bibfnamefont
  {R.}~\bibnamefont {Sabatini}}, \bibinfo {author} {\bibfnamefont
  {B.}~\bibnamefont {Santra}}, \bibinfo {author} {\bibfnamefont
  {M.}~\bibnamefont {Schlipf}}, \bibinfo {author} {\bibfnamefont {A.~P.}\
  \bibnamefont {Seitsonen}}, \bibinfo {author} {\bibfnamefont {A.}~\bibnamefont
  {Smogunov}}, \bibinfo {author} {\bibfnamefont {I.}~\bibnamefont {Timrov}},
  \bibinfo {author} {\bibfnamefont {T.}~\bibnamefont {Thonhauser}}, \bibinfo
  {author} {\bibfnamefont {P.}~\bibnamefont {Umari}}, \bibinfo {author}
  {\bibfnamefont {N.}~\bibnamefont {Vast}}, \bibinfo {author} {\bibfnamefont
  {X.}~\bibnamefont {Wu}}, \ and\ \bibinfo {author} {\bibfnamefont
  {S.}~\bibnamefont {Baroni}},\ }\href {\doibase 10.1088/1361-648x/aa8f79}
  {\bibfield  {journal} {\bibinfo  {journal} {Journal of Physics: Condensed
  Matter}\ }\textbf {\bibinfo {volume} {29}},\ \bibinfo {pages} {465901}
  (\bibinfo {year} {2017})}\BibitemShut {NoStop}%
\bibitem [{\citenamefont {Belenkov}(2001)}]{Belenkov2001}%
  \BibitemOpen
  \bibfield  {author} {\bibinfo {author} {\bibfnamefont {E.~A.}\ \bibnamefont
  {Belenkov}},\ }\href {\doibase 10.1023/a:1011601915600} {\bibfield  {journal}
  {\bibinfo  {journal} {Inorganic Materials}\ }\textbf {\bibinfo {volume}
  {37}},\ \bibinfo {pages} {928} (\bibinfo {year} {2001})}\BibitemShut
  {NoStop}%
\end{thebibliography}%

\end{document}